\documentclass[reprint,amsmath,amssymb,aps,superscriptaddress,twocolumn,nofootinbib,10pt]{revtex4-1}
\usepackage[colorlinks=true,allcolors=blue]{hyperref}
\usepackage{graphicx,color}
\usepackage{dcolumn}
\usepackage{subfigure}
\usepackage{bm}
\usepackage{amsmath,amsfonts,amssymb}
\usepackage{acronym}
\usepackage{enumitem}
\usepackage{array}
\usepackage{float} 
\usepackage{multirow}
\usepackage{CJK}
\allowdisplaybreaks
\newcommand{\be}{\begin{equation}}
\newcommand{\ee}{\end{equation}}
\newcommand{\bea}{\begin{eqnarray}}
\newcommand{\eea}{\end{eqnarray}}

\newcommand{\SOUTHCUTa}{Department of physics, Nanchang University, Nanchang, 330031, China}
\newcommand{\SOUTHCUTb}{Center for Relativistic Astrophysics and High Energy Physics, Nanchang University, Nanchang, 330031, China}
\newcommand{\IIT}{Indian Institute of Technology, Gandhinagar, Gujarat-382355, India}

\newacro{EMRI}{extreme mass-ratio inspirals}
\newacro{MBH}{massive black hole}
\newacro{BH}{black hole}
\newacro{GR}{general relativity}
\newacro{HKBH}{hairy Kerr black hole}
\newacro{KNBH}{Kerr-Newmann black hole}
\newacro{KBH}{Kerr black hole}
\newacro{NHT}{no-hair theorem}
\newacro{DWD}{double white dwarf}
\newacro{GW}{gravitational wave}
\newacro{AK}{analytic kludge}
\newacro{NK}{numerical kludge}
\newacro{AAK}{augmented analytic kludge}
\newacro{CO}{compact object}
\newacro{PE}{parameter estimation}
\newacro{SNR}{signal-to-noise ratio}
\newacro{PN}{post newtonion}
\newacro{FIM}{Fisher information matrix}
\newacro{LSO}{last stable orbit}
\newacro{ISCO}{innermost stable circular orbit}
\newacro{BBH}{Binary Black Hole}
\newacro{BNS}{Binary Neutron Star}
\newacro{NS}{Neutron Star}
\newacro{KN}{Kerr-Newmann}
\newcommand{\beq}{\begin{equation}}
\newcommand{\eeq}{\end{equation}}
\newcommand{\beqa}{\begin{eqnarray}}
\newcommand{\eeqa}{\end{eqnarray}}

\def\lsim{\mathrel{\rlap{\lower4pt\hbox{\hskip0.5pt$\sim$}}
    \raise1pt\hbox{$<$}}}         
\def\gsim{\mathrel{\rlap{\lower4pt\hbox{\hskip0.5pt$\sim$}}
    \raise1pt\hbox{$>$}}}         
\begin{document}
\begin{CJK*}{UTF8}{gbsn}
\title{{\Large {\bf 
Probing scalar field with generic extreme mass-ratio inspirals around Kerr black holes}}}

\author{Tieguang Zi}
\email{zitieguang@ncu.edu.cn}
\affiliation{\SOUTHCUTa}
\affiliation{\SOUTHCUTb}

\author{Shailesh Kumar}
\email{shailesh.k@iitgn.ac.in} \email{shaileshkumar.1770@gmail.com}
\affiliation{\IIT}

\begin{abstract}
The future space-based gravitational wave observatories are expected to provide unprecedented opportunities to explore intricate characteristics of black hole binaries, particularly for extreme mass-ratio inspirals (EMRIs), in which a stellar-mass compact object slowly inspirals into a supermassive black hole. These systems are very prominent sources for testing gravity in the strong gravity fields and for probing potential deviations from general relativity, including those arising from the presence of fundamental scalar fields. In this work, we examine the impact of a scalar charge carried by the inspiraling object within the context of EMRIs. We focus on generic orbits that present both eccentricity and inclination to evaluate how these parameters affect the modifications induced by the scalar charge to the gravitational wave signal. Our results demonstrate that the inclusion of orbital inclination, in particular, enhances the detectability of scalar field effects by introducing richer waveform features that deviate from the purely general relativistic case. The interplay among scalar charge, eccentricity and inclination provides a more complete sampling of the black hole spacetime, suggesting that EMRIs with such generic orbits represent compelling systems for stringently constraining or discovering new fundamental fields through future gravitational wave observations.

\end{abstract}
\maketitle
\end{CJK*}
\section{Introduction}
The direct detection of gravitational waves (GWs) has opened several new frontiers in testing the fundamental aspects of gravity and the foundations of general relativity (GR) in strong-field regime. After three observing runs, the LIGO-Virgo-KAGRA collaborations have detected the catalogs of about 90 events of compact object mergers, including the upcoming fourth observational runs \cite{KAGRA:2013rdx}, the total events will reach up to $\sim200$ GW signals, allowing us to test GR and search for deviations \cite{LIGOScientific:2016aoc,LIGOScientific:2017vwq,KAGRA:2021vkt,LIGOScientific:2021sio,Xie:2024xex,Berti:2025hly,Yunes:2025xwp}. 
The next generation experiments, such as Einstein Telescope (ET) \cite{Punturo:2010zz}, Laser Interferometer Space Antenna (LISA) \cite{LISA:2017pwj}, TianQin \cite{TianQin:2015yph, TianQin:2020hid} and Taiji \cite{Ruan:2018tsw}, will be equipped with a higher sensitivity and aim to detect several binary populations with a bigger signal-to-noise ratio (SNR) in the lower frequency bands. Among these detectors,  space-based detectors have been expected to explore fundamental physics with unprecedented precision. The events observed to date by ground-based GW observatories have focused on nearly equal-mass binaries; however, the future space detectors will access lower-frequency bands where small mass-ratio binaries, particularly ``extreme mass-ratio inspirals'' (EMRIs) are expected to be one of the prominent sources for testing GR and beyond. These systems\textemdash comprising a stellar-mass compact object (secondary with mass $m_p$) slowly spiralling into a supermassive black hole (SMBH, primary with mass $M$), maintaining a mass-ratio in the range ($q\equiv m_{p}/M = 10^{-7} - 10^{-4}$)\textemdash produce long-duration, complex waveforms that encode detailed information about the background spacetime. We know that numerical relativity (NR) has played a key role in advancing waveform modeling of binaries \cite{Pretorius:2005gq, Campanelli:2005dd, Buonanno:2006ui}; however, its direct implementation for EMRIs sources becomes challenging. The computational demands and requirements grow significantly as the mass asymmetry increases, making black hole (BH) perturbation techniques more approachable for modeling EMRIs. Since EMRIs exhibit a large number of orbital cycles, from thousands to millions, around a central massive BH, they offer an unparalleled opportunity to map spacetime geometries with exquisite accuracy. Their sensitivity to subtle gravitational effects makes them promising probes of deviations from GR, including the presence of additional fields \cite{Berry:2019wgg,LISA:2022yao,Barack:2018yly,Barack:2006pq, Babak:2017tow, Fan:2020zhy, Zi:2021pdp,LISA:2022kgy}. 

We know that compact binary systems emit gravitational radiation primarily through tensorial polarizations in GR, with the leading-order contribution governed by the mass quadrupole moment, consequently, the quadrupole formula has been successful in explaining observations, particularly for systems involving BHs and neutron stars. However, numerous non-GR theories of gravity\textemdash such as Brans-Dicke and other scalar-tensor\textemdash predict the existence of additional gravitational degrees of freedom. These theories often introduce scalar fields that are coupled to matter and curvature, allowing the emission of dipolar radiation components that are absent in pure GR \cite{Alsing:2011er, Saijo:1996iz}. Such modes can potentially provide interesting signatures of new gravitational physics; however, they are generally much weaker than the standard tensor modes in the low-frequency regime, particularly in the milli-Hertz range targeted by future space-based GW observatories \cite{dePaula:2004bc, Hou:2017bqj}. Nevertheless, their cumulative effects over long inspiral durations can leave detectable footprints on EMRIs. Even though detection prospects of such scalar or vector modes remain highly challenging due to their suppressed amplitudes; however, building on this, researchers have explored how such scalar effects could be observed without any specific gravity model and have developed general waveform templates to search for new fundamental fields \cite{Maselli:2020zgv, Maselli:2021men}. More recent efforts have expanded the analysis to more realistic settings, including BHs with spin, eccentric orbits, and cases where the scalar field has mass \cite{Guo:2022euk, Barsanti:2022ana, Barsanti:2022vvl, Zhang:2022rfr, Zhang:2022hbt}. 

In this line of endeavour, EMRIs offer a unique observational window into such phenomena and are expected to detect the new signatures at the horizon scale, multipolar structure of primary objects,  existence of exotic compact object and new fundamental fields coupled to gravity theories. As mentioned, many such theories introduce new degrees of freedom\textemdash especially scalar fields\textemdash which can be coupled to curvature or matter and modify the dynamics of compact objects in interesting ways \cite{Saijo:1996iz, Alsing:2011er, Kuntz:2020yow}. EMRIs are particularly well suited to detect these effects, as the cumulative influence of such fields can give observable signatures on the orbital motion and emitted waveforms. Eccentric orbits are particularly important in EMRIs systems when searching for signatures of scalar fields or other non-GR effects, as such systems can probe a wider as well as more dynamic region of the spacetime around the primary. Unlike circular orbits, eccentric trajectories bring the secondary object much closer to the BH at periapsis, where spacetime curvature and any potential coupling to additional fields are strongest. Many recent studies have considered numerous observable effects through eccentric EMRIs based on Teukolsky approach, metric perturbation and post-Newtonian (PN) techniques \cite{Barsanti:2022ana, Duque:2024mfw, Chapman-Bird:2025xtd, Munna:2020som, Zhang:2020rxy, Zi:2022hcc, Zi:2023qfk, Kumar:2025jsi, Kumar:2025njz, Kumar:2024our, Zi:2024jla, AbhishekChowdhuri:2023gvu, Kumar:2024utz, Rahman:2023sof}. Therefore, eccentric motion with generic orbits (non-equatorial) will further enhance the sensitivity of the system such as scalar radiation. Moreover, eccentric orbits generate a richer GW spectrum with higher harmonics, offering more ways through which non-GR effects can be more prominent for the detectability. These features make eccentric EMRIs powerful probes for detecting even subtle imprints of new fields and for testing gravity in the strong-field regime. With this motivation, the present article focuses on hunting the signatures of scalar charge with inclined eccentric orbits, in other words, generic eccentric orbits from the future LISA observations.

Let us briefly look at how the draft is organized. In section (\ref{BH}), we provide a theoretical setup of the problem that describes the general action with scalar field and the perturbation equations. In section (\ref{geodesic1}), we structure the geodesic equations and fundamental frequencies for the Kerr BH. Section (\ref{inspiral}) serves the details for the gravitational and scalar fluxes under adiabatic evolution, including the waveform details in section (\ref{waveform&mismatch}). Further, we provide the results of fluxes, dephasing and mismatch in order to examine the prospects of determining the signatures of the scalar charge with LISA observations in section (\ref{wave}). Finally, we discuss the final remarks of the investigation in section (\ref{dscn}) with future prospets.

\section{Theoretical setup}\label{BH}
In order to describe scalar emission in EMRIs, we start with a general scalar-tensor action in which a  massless real scalar field $\psi$ interacts with gravity through non-minimal couplings \cite{Ramazanoglu:2016kul, Maselli:2020zgv, Maselli:2021men, Barsanti:2022vvl, Spiers:2023cva, Zi:2024mbd}. We begin by considering the general action, given as follows:
\begin{align}\label{actions}
S[\mathbf{g},\psi,\Psi] =  S_0[\mathbf{g},\psi]+ \alpha S_c[\mathbf{g}, \psi] + S_m[\mathbf{g},\psi,\Psi],
\end{align}
where $\mathbf{g}$ is the spacetime metric, $\psi$ is the massless scalar field and 
\begin{equation}
 S_0[\mathbf{g},\psi] = \int d^4x \frac{\sqrt{-g}}{16\pi} \left(R-\frac{1}{2}\partial_\mu\psi \partial^\mu \psi\right),
\end{equation}
with $R$ being the Ricci scalar. The action $\alpha S_c$ denotes the coupling between the scalar field and the metric, where the constant $\alpha$ has a dimension of $[\rm mass]^n$ with $n\geq1$ and its effect is governed by a characteristic energy scale. The action of the matter field $\Psi$ is determined by $S_m[\mathbf{g},\psi,\Psi]$.
According to the action \eqref{actions}, one can obtain the following two field equations,
\begin{equation}
\begin{aligned}
G_{\mu\nu} =& 8\pi T_{\mu\nu}^{\rm scal} + \alpha T^c_{\mu\nu}  +  T^p_{\mu\nu} + T_{\mu\nu}^m, \\ \nabla_\mu \nabla^\mu \psi =&  T^c  +  T^m,
\end{aligned}
\end{equation}
where 
\begin{equation}
\begin{aligned}
T_{\mu\nu}^{\rm scal} & = \frac{1}{16\pi} \left[\partial_\mu\psi \partial^\mu \psi - \frac{1}{2}g_{\mu\nu} (\partial \psi)^2\right]\;, \nonumber\\
T^c_{\mu\nu} & =  -\frac{16\pi}{\sqrt{-g}} \frac{\delta S_c}{\delta g^{\mu\nu}}\;, \nonumber\\
T^m & =  -\frac{16\pi}{\sqrt{-g}} \frac{\delta S_m}{\delta \psi}\;.
\end{aligned}
\end{equation}
When the constant $\alpha = 0$, the scalar-tensor theories hold the no-hair theorem~\cite{Bekenstein:1995un,Hawking:1972qk,Sotiriou:2011dz,Hui:2012qt}. The magnitude of the deviation from the GR results depends on a dimensionless parameter $\xi$ termed as
\begin{align}\label{eq:xi}
\xi = \frac{\alpha}{M^n} =q^n \frac{\alpha}{\mu^n} \equiv q^{n} \xi_p\;,
\end{align}
where $q=\mu/M$ is the mass-ratio of binaries with $\mu$ being the mass of the secondary and $M$ being the mass of the central supermassive primary. We keep $q \ll 1$ for our present analysis relevant for EMRIs. The departures from the Kerr geometry are controlled by the dimensionless coupling $\xi$. Imposing $\xi_{p} < 1$, as required by existing astrophysical bounds \cite{Nair:2019iur}, ensures that $\xi \ll 1$. In EMRIs with supermassive primaries, this parameter is therefore exceedingly small, so the exterior spacetime is well approximated by the vacuum Kerr solution.


In the perturbative method, for the metric and scalar field equations, EMRIs can be described by expanding the powers of mass-ratio $q$
\begin{align}\label{eq:gmunu}
g_{\mu\nu} = g_{\mu\nu}^{(0)} + q h_{\mu\nu}^{(1)}\;, 
\psi = \psi^{(0)} + q \psi^{(1)}\;.
\end{align}
In this paper, we only focus on the leading  dissipative contribution determined by the first order perturbations $h_{\mu\nu}^{(1)}$ and $\psi^{(1)}$. At the leading-order in the mass-ratio, the no-hair theorem~\cite{Chase:1970omy,Bekenstein:1995un,Hawking:1972qk,Hui:2012qt} implies that the scalar field admits only a constant solution, which we set to zero without loss of generality. Consequently, the spacetime background of the massive black hole reduces to the Kerr geometry.
On the other hand, considering the first order perturbations,
the scalar and metric equations are sourced by a moving secondary described by the point-particle with a mass.
In the skeletonized approach~\cite{Eardley:1975fgi,TDamour_1992}, the matter field action $S_m$ can be replaced with the point-particle action $S_p$. In other words, to model the secondary body in the EMRI, we adopt an effective point-particle description appropriate for compact objects, where the full matter action associated with the secondary is replaced by a skeletonized worldline action that captures its interaction with the gravitational and scalar fields. Thus, the secondary object carrying a scalar charge can be described by the following action \cite{Maselli:2020zgv, Eardley:1975fgi, TDamour_1992}
\begin{equation}
S_p = -\int m(\psi) \sqrt{g_{\mu\nu} \frac{dx^\mu}{d\tau} \frac{dx^\nu}{d\tau} d\tau},
\end{equation}
where $\frac{dx^\nu}{d\tau}$ is the four-velocity and $\tau$ is the proper time. The mass function $m(\psi)$ can be computed with the value of the scalar field at the location of the particle, which is obtained by the features of moving secondaries.

Assuming that the center of the inspiraling object has a reference frame $\{\tilde{x}_\mu\}$, the solution of the scalar field equation in the world-tube of the secondary,  in which the background can be the flat spacetime perturbation, can be written as
\begin{equation}\label{eq:psi:expanision}
\psi = \psi_0 + \frac{\mu q_s}{\tilde{r}} + \mathcal{O}\left(\frac{\mu^{2}}{\tilde{r}^2}\right),
\end{equation}
where $q_s$ is the dimensionless scalar charge carried by the secondary and $\tilde{r}$ is the distance from the worldline satisfying the relation $m_p\ll \tilde{r} \ll M$. Substituting Eq.~\eqref{eq:psi:expanision} into the scalar field equation and matching the solution in the buffer region, one can obtain the relation of scalar charge and mass function $m'(\psi_0) = -\mu q_s/4$ and $m(\psi_0)=m_p$, and a detailed demonstration can refer to Ref.~\cite{Maselli:2021men,Barsanti:2022ana}.

On the other hand, the contributions of the quantities $T_{\mu\nu}^{\rm scal}$ and $\alpha T_{\mu\nu}^{\rm c}$ are embodied at the higher order of mass-ratio expansions ($\mathcal{O}(q^2)$), so they can be ignored for the leading order expansion ($\mathcal{O}(q)$). For EMRIs, action $S_c$ should be computed on the background of MBH, which have a dimension of primary's mass $M$. Moreover, action is also dimensions, $[S_0] = (\rm mass)^2$ and $[S_c]=(\rm mass)^{2-n}$,
so one can deduce that 
\begin{equation}
S_c \sim M^{-n}S_0
\end{equation}
and
\begin{equation}
\alpha T^c_{\mu\nu} = -\frac{16\pi\alpha}{\sqrt{-g}} \frac{\delta S_c}{\delta g^{\mu\nu}}\sim  -\frac{16\pi \alpha M^{-n}}{\sqrt{-g}} \frac{\delta S_0}{\delta g^{\mu\nu}}\;,
\end{equation}
and 
\begin{equation}
\alpha T^c_{\mu\nu} \sim \xi G_{\mu\nu}\ll G_{\mu\nu}
\end{equation}
due to $\xi\equiv\alpha M^{-n} \ll1$.
Similarly, for the scalar field equation, we still obtain the approximation relation
\begin{equation}
-\frac{16\pi\alpha}{\sqrt{-g}} \frac{\delta S_c}{\delta \psi}\sim  -\frac{16\pi \alpha M^{-n}}{\sqrt{-g}} \frac{\delta S_0}{\delta \psi} \sim \xi \nabla_\mu \nabla^\mu \psi \ll 
\nabla_\mu \nabla^\mu \psi\;.
\end{equation}
Therefore, we finally get the scalar and metric field equations at the first order of the mass-ratio
\begin{eqnarray}
G^{\mu\nu}[h^{(1)}_{\mu\nu}]  = 8\pi \mu \int \frac{\delta^{(4)} (x-x(\tau))}{\sqrt{-g}} \frac{dx^\mu}{d\tau}\frac{dx^\nu}{d\tau} d\tau \label{eq:grav:pert}\;,\\
\nabla_\mu \nabla^\mu \psi^{(1)} = -4\pi q_s \mu \int \frac{\delta^{(4)} (x-x(\tau))}{\sqrt{-g}} d\tau  \label{eq:scal:pert}\;,
\end{eqnarray}
where the superscript $\text{``(1)"}$ denotes the linearized perturbations of the fields. We omit writing this superscript further just for convenience of writing. In essence, the gravitational field equation in Eq.~\eqref{eq:grav:pert} is consistent with that of the GR case. The scalar field equation Eq.~\eqref{eq:scal:pert} has an extra source term, where the scalar charge endowed by the secondary controls the magnitude of the scalar perturbation. The scalar charge $q_s$ determines the deviation of the evolution of the EMRI beyond GR. In fact, the scalar field equation can be mapped to some parameters in most modified gravity theories. Thus, the tighter constraint of the scalar charge $q_s$ with future LISA detection would be re-incorporated into the measurements of parameters relating to modified theories \cite{Maselli:2021men,Julie:2022huo,Speri:2024qak}.

\section{Generic timelike geodesics in Kerr spacetime} \label{geodesic1}
The geodesic equations of the Kerr BH in Boyer-Lindquist coordinates $(t,r,\theta, \phi)$ can be written as
\begin{eqnarray}\label{geodesic}
\Sigma\frac{dt}{d\tau} &=& 
V_{t \theta}(r,\theta) + V_{t r}(r,\theta) 
\nonumber \\
&=& E \Big[\frac{\left(a^2+r^2\right)^2}{\Delta} - a^2\sin^2\theta \Big] \nonumber \\ &+& aL_z\left(1-\frac{a^2+r^2}{\Delta}\right) \label{eq:geo:dt}\,, \\
\Sigma\frac{d\phi}{d\tau} &=&  V_{\phi \theta}(r,\theta) + V_{\phi r}(r,\theta)
\nonumber\\ &=&
\frac{L^2}{\sin^2\theta}-\frac{a^2L}{\Delta} +a E\left( \frac{a^2+r^2}{\Delta} -1\right) \label{eq:geo:dphi}\;, \\
\Sigma^{2}\Big(\frac{dr}{d\tau}\Big)^{2} &=& R(r) = \Big(E(a^{2}+r^{2})-aL\Big)^{2} 
\nonumber\\ &-&\Delta\Big[r^{2}+Q+(L-aE)^2\Big] \label{eq:geo:dr}\,, \\
\Sigma^{2}\Big(\frac{d\theta}{d\tau}\Big)^{2} &= & \Theta(\theta) \nonumber\\ &=& Q - \cos^{2}\theta \Big(a^2(1-E^2)-\frac{L^2}{\sin^2\theta}\Big) \label{eq:geo:dz}\;,
\end{eqnarray}
where $E$ and $L$ are the orbital energy and angular momentum of a moving particle per unit mass $\mu$, $Q$ is the Carter constant per unit mass square $\mu^2$ and a function $\Delta=a^2+r^2-2Mr$ associated with the Kerr metric.
For a bound geodesic orbit in the Kerr spacetime, the orbital energy is confined in the range of $0\leq E\leq1$, the Carter constant is non-negative $(Q\geq0)$. The Mino time 
$d/d\lambda= \Sigma d/d\tau$ is useful in the computation of orbital fundamental frequencies; the equations are rewritten as the familiar forms~\cite{Schmidt:2002qk}.

For the radial motion, the radial function $R(r)$ varies between two turning points termed the apastron $r_a$ and the periastron $r_p$, which satisfies the condition $R(r_a=p/(1-e))=R(r_p=p/(1+e))=0$. The radial coordinate can be expressed with an angular parameter $\varphi$: 
\begin{equation}\label{eq:rpsi}
r(\varphi) = \frac{p M}{1+e\cos\varphi}\;,
\end{equation}
where $\varphi$ changes from $\varphi=0$ (periastron) to $\varphi=\pi$ (apastron). For the polar motion, the angular polar can be parametrized in terms of a new angular variable $\chi$ 
\begin{equation}\label{eq:zchi}
 z\equiv \cos^2\theta  = z_- \cos^2\chi\;, 
\end{equation}
where $z_-$ is one of the roots of the quadratic equation $\Theta(\theta_{\min, \max})=0$ and $\theta_{\min, \max}$ is the minimum and maximum value during the angle $\theta$ varies periodically.
It is convenient to introduce an angle of inclination $I\equiv\pi/2-{\rm sgn} (L)\theta_{\min}$, it changes from zero to $\pi$ for the prograde motion in the equatorial plane,
the parameter $x=\cos I$ describes well the orbital inclination and varies in the range of $[1,-1]$ when the secondary is motion from prograde to retrograde in the equatorial plane. So, the parameter $(p,e,x)$ that describes the generic geodesics is equivalent to the integrals of motion $(E,L,Q)$, which is also convenient to evolve adiabatically inspiraling geodesic orbits. A good scheme is to obtain $(E,L,Q)$ of the geodesic equations for Kerr spacetime, argued by Schmidt and van de Meent \cite{Schmidt:2002qk,vandeMeent:2019cam}, in which the integrals of the generic orbital motion can be expressed in terms of the orbital parameters $(p,e,x)$.

Since the radial and polar solutions are periodic in Mino time, they both have orbital frequencies, $\Upsilon_r$ and $\Upsilon_\theta$. Using the parameterized equations in Eqs. \eqref{eq:rpsi} and \eqref{eq:zchi}, one can rewrite the geodesic equations to evaluate the coordinate time frequencies, where the duration of Mino time varies at the radial and polar period is constant $\Gamma$. These quantities can be used to compute the orbital fundamental frequencies 
\begin{eqnarray}
\Omega_r = \frac{\Upsilon_r}{\Gamma}~,
\Omega_\theta = \frac{\Upsilon_\theta}{\Gamma}~,
\Omega_\phi = \frac{\Upsilon_\phi}{\Gamma}~.
\end{eqnarray}
A detailed and complete argument for computing these frequencies has been shown in ~\cite{Schmidt:2002qk,Drasco:2005kz,Fujita:2009bp,vandeMeent:2019cam}. The total orbital frequency is made up of the linear combination of  three fundamental frequencies
\begin{equation}
\omega\equiv\omega_{mkn} = m\Omega_\phi + n \Omega_r + k \Omega_\theta\;,
\end{equation}
where parameters $(m,k,n)$ are integers. \vspace{0.5cm}

\section{Adiabatic evolution of EMRIs orbits} \label{inspiral}
\subsection{Gravitational fluxes}\label{grav:flux}
In this section,  we introduce the scalar and gravitational perturbations of Kerr BH derived by Teuokolsky formalism~\cite{Teukolsky:1973ha,Teukolsky:1972my}.
We can write down the separable radial and angular equations in the following fashion
\begin{widetext}
\begin{eqnarray}
\begin{aligned}
\Delta^{-s}\frac{d}{dr}\Big(\Delta^{s+1}\frac{dR_{s}}{dr} \Big) + \Big[\frac{K(K-is\Delta')}{\Delta}+4is\omega r-\lambda \Big]& R_{s} = \mathcal{T}_{s\ell m\omega} \;,  \label{eq:radial:teuko} \\
\Big[\frac{1}{\sin\theta}\frac{d}{d\theta}\Big(\sin\theta\frac{d}{d\theta}\Big)-\Big(\frac{m+s\cos\theta}{\sin\theta}\Big)^{2}+s+\lambda_{s}\Big]& {}_{s}S_{\ell m} = 0 \;, \label{eq:angular:teuko}
\end{aligned}
\end{eqnarray}
\end{widetext}
where $K=(r^{2}+a^2)\omega -am$ and $s\in (0, 1, \pm 2)$.
The equation for the $s=-2$ case denotes the gravitational perturbation, and the $s=0$ case is the scalar perturbation. $\lambda_{s} = (\ell-s)(\ell+s+1)$ is the eigenvalue of the spin-weighted spherical harmonics (${}_{s}S_{\ell m}$). $\mathcal{T}_{slm\omega}$ is the source term of the inspiraling object, described in appendix (\ref{app:source:grav}) and $R_{s}(r)$ is the radial function.
For the gravitational radial equation, its homogeneous solution obeys the following asymptotic behavior
\begin{equation}
\begin{aligned}\label{rty1}
R^{in}_{\ell m\omega}(r)\sim & \begin{cases}
    B^{tran}_{\ell m\omega}\Delta^{2}e^{-i\mathcal{K}_\omega r_{*}}; & \text{$r\longrightarrow  r_{+}$},\\
    B^{out}_{\ell m\omega}r^{3}e^{i\omega r_{*}}+B^{in}_{\ell m\omega}r^{-1}e^{-i\omega r_{*}}; & \text{$r\longrightarrow \infty$}, 
  \end{cases}\\
R^{up}_{\ell m\omega}(r)\sim & \begin{cases}
    C^{out}_{\ell m\omega}e^{i \mathcal{K}_\omega r_{*}}+C^{in}_{\ell m\omega}\Delta^{2}e^{-i\mathcal{K}_\omega r_{*}}, & \text{$r\longrightarrow r_{+}$},\\
    C^{tran}_{\ell m\omega}r^{3}e^{i\omega r_{*}}; & \text{$r\longrightarrow \infty$},
  \end{cases}
  \end{aligned}
\end{equation}
where $\mathcal{K}_\omega=\omega_{mkn}-am/2r_+$, $r_+=M+\sqrt{M^2-a^2}$ and
$r^\ast$ is the tortoise coordinate defined by $dr^\ast/dr = (r^2+a^2)/\Delta$~\cite{Teukolsky:1973ha}.
Since the radial equation has a long-range potential, one can convert it into the Sasaki-Nakamura (SN) equation with a short-range potential, enabling for an efficient numerical solution~\cite{PhysRevD.102.024041, 10.1143/PTP.67.1788, Kumar:2025njz}.
The general solutions of the radial equation are combined with two homogeneous solutions \eqref{rty1}
\begin{equation}
R_{\ell m\omega}(r) = R^{in}_{\ell m\omega}(r) Z^{\infty}_{\ell m\omega}(r) +  R^{up}_{\ell m\omega}(r) Z^{H}_{\ell m\omega}(r), 
\end{equation}
where 
\begin{align}\label{ampl:grav}
Z^{H}_{\ell m\omega} =& \frac{1}{2i\omega B^{in}_{\ell m\omega}} \int_{r_{+}}^{r}\frac{R^{in}_{\ell m\omega}}{\Delta^{2}}\mathcal{T}_{\ell m\omega} dr\,, \\ Z^{\infty}_{\ell m\omega} =& \frac{B^{tran}_{\ell m\omega}}{2i\omega B^{in}_{\ell m\omega}C^{tran}_{\ell m\omega}}\int_{r}^{\infty}\frac{R^{up}_{\ell m\omega}}{\Delta^{2}}\mathcal{T}_{\ell m\omega} dr\,.
\end{align}
The amplitudes $Z^{H,\infty}_{\ell m\omega}$ are useful for calculating the fluxes at horizon and infinity, which involve the integration over 
the source term sourced by the secondary using two homogeneous solutions. The explicit expressions of
the source terms for the various orbital configurations have been argued and verified broadly in ~\cite{Sago:2005fn,Drasco:2005kz,Isoyama:2018sib,Hughes:2021exa}. We also briefly review the process of computing two amplitudes in the appendix~(\ref{app:source:grav}) and (\ref{app:source:scal}).
Note that the ingoing and outgoing homogeneous solutions 
$(R_{\ell m\omega}^{in}, R_{\ell m\omega}^{up})$ are obtained by imposing the ingoing boundary condition near the horizon and the outgoing condition at the infinity, which can be solved with the method proposed by Mano, Suzuki and Takasug $(\texttt{MST})$ and the series expanding method by Jiang-Han \cite{Mano:1996vt,Jiang:2025mna}. In this paper, we adopt the analytic homogeneous solutions based on  $\texttt{MST}$ method. For a comprehensive introduction of this, we refer the readers to the classic works~\cite{Hughes:1999bq,Drasco:2005kz,Hughes:2021exa}.

For an EMRI system, the changing rate of the energy, angular momentum and Carter constant is termed fluxes, emitting to infinity and absorbed by the horizon, which can be estimated as ~\cite{Sago:2005fn,Isoyama:2018sib,Hughes:2021exa}
\begin{equation}\label{fluxes:grav:eqs}
\begin{aligned}
\dot{E}_G^\infty &\equiv\Big(\frac{dE}{dt}\Big)^{\infty}_{\rm Grav}  = \sum_{\ell mkn} \frac{|Z^{H}_{\ell mkn}|^2}{4\pi \omega_{mkn}^2} \\
\dot{E}_G^H&\equiv\Big(\frac{dE}{dt}\Big)^{H}_{\rm Grav} = \sum_{\ell mkn}\alpha_{\ell mkn} \frac{|Z^{\infty}_{lmkn}|^2}{4\pi \omega_{mkn}^2}, \\
\dot{L}_G^\infty& \equiv \Big(\frac{dL}{dt}\Big)^{\infty}_{\rm Grav}  = \sum_{\ell mkn} \frac{m |Z^{H}_{\ell mkn}|^2}{4\pi \omega_{mkn}^3};\\
\dot{L}_G^H &\equiv\Big(\frac{dL_z}{dt}\Big)^{H}_{\rm Grav} = \sum_{\ell mn} \frac{\alpha_{\ell mkn} m|Z^{\infty}_{\ell mkn}|^2}{4\pi \omega_{ mkn}^3}\;,\\
\dot{Q}_G^H &\equiv\Big(\frac{dQ}{dt}\Big)^{H}_{\rm Grav} = \sum_{\ell mkn} \frac{\alpha_{\ell mkn} |Z^{H}_{\ell mkn}|^2}{2\pi \omega_{mkn}^3}\left(\mathcal{L}_{\ell mkn}+k\Upsilon_\theta\right)\;,\\
\dot{Q}_G^\infty&\equiv\Big(\frac{dQ}{dt}\Big)^{\infty}_{\rm Grav} = \sum_{\ell mkn} \frac{|Z^{\infty}_{\ell mkn}|^2}{2\pi \omega_{mkn}^3}\left(\mathcal{L}_{\ell mkn}+k\Upsilon_\theta\right)\;\\
\end{aligned}
\end{equation}
The coefficients $(\alpha_{\ell mkn},\mathcal{L}_{\ell mkn})$ are easily
obtained with the orbital parameters and frequencies, which are indeed lengthy; their full expressions can be found in ~\cite{Sago:2005fn,Drasco:2005kz,Hughes:2021exa}.
Therefore, the total changing rate of the integrals of motions for the Kerr geodesics are the sum of fluxes at infinity and horizon as follows: 
\begin{equation}
\dot{\mathcal{C}}_{G} = \dot{\mathcal{C}}^\infty_{G} + \dot{\mathcal{C}}_{G}^H~~,
\end{equation}
 where $\mathcal{C}\in (E, L, Q)$. 
 When computing the fluxes over the modes $(\ell,m,n,k)$, we numerically integrate over the angels $(\chi,\varphi)$ that appear from the parameterization \eqref{eq:rpsi} and \eqref{eq:zchi}. This implies the evaluation of fluxes at different orbital radii. 
\subsection{Scalar fluxes}\label{scal:flux}
The scalar equation for $s=0$ in Eq.~\eqref{eq:radial:teuko}
can be simplified as 
\begin{equation}\label{eq:scal:radial}
\frac{d}{dr}\left( \Delta \frac{ d}{dr}R^s_{\ell m\omega} (r,\omega)\right) + V(r)R^s_{\ell m\omega} (r,\omega) 
= T^s_{\ell m\omega}(r, \omega)  
\end{equation}
where the scalar potential function is $V(r) = K^2/\Delta - \lambda$ and the source $T^{s}_{\ell m\omega}$ can be found in the appendix (\ref{app:source:scal}). We adopt the Green's function to compute the homogeneous solutions, then obtain the scalar fluxes by integrating the solutions over the source terms.
The solutions to equation \eqref{eq:scal:radial} have the following  boundaries
\begin{equation}
\begin{aligned}\label{scal:asyms}
R^{s,-}_{\ell m\omega}(r)\sim & \begin{cases}
 e^{-i\mathcal{K}_\omega r_{*}}; & \text{$r\longrightarrow  r_{+}$},\\
    A^{out}_{\ell m\omega} r^{-1} e^{i\omega r_{*}}
    +A^{in}_{\ell m\omega}r^{-1}e^{-i\omega r_{*}}; & \text{$r\longrightarrow \infty$}, 
  \end{cases}\\
R^{s,+}_{\ell m\omega}(r)\sim & \begin{cases}
   B^{out}_{\ell m\omega}e^{i \mathcal{K}_\omega  r_{*}}
    +B^{in}_{\ell m\omega}e^{-i\mathcal{K}_\omega r_{*}}, & \text{$r\longrightarrow r_{+}$},\\
    e^{i\omega r_{*}}; & \text{$r\longrightarrow \infty$} \;,
  \end{cases}
  \end{aligned}
\end{equation}
One can see that the boundaries in Eq.~\eqref{scal:asyms} including the terms of $r^{-1}$ have a poor convergence, 
following the method in Refs.~\cite{Maselli:2021men,Barsanti:2022ana,DellaRocca:2024pnm},
one can recast the quantity $R^s_{\ell m\omega} = \psi^s_{\ell m\omega}/\sqrt{a^2+r^2}$ to avoid this problem. The solution $\psi^s$ obeys the pure ingoing asymptotic behaviour at the horizon and the pure outgoing behavior at the infinity. The general solution at the horizon and infinity is integrated
over the source term
\begin{equation}\label{ampls:scal}
Z_{\ell mkn}^{s,\pm} = \int_{-\infty}^\infty \frac{\Delta}{(r^2+a^2)^{3/2}} \frac{T^s_{\ell m\omega}(r',\omega) 
\psi^{s,\mp}_{\ell m\omega} }{W} dr'\;,
\end{equation}
where $W=\psi_{\ell m\omega}'^{s,+} \psi^{s,-}_{\ell m\omega} - \psi_{\ell m\omega}'^{s,-} \psi^{s,+}_{\ell m\omega}$ is the Wronskian and the prime is the time derivative with respect to the tortoise coordinate. The quantities $Z_{\ell mkn}^{s,\pm}$ are called the amplitudes for scalar emission at the horizon and infinity, which shows how to compute the numerical integration for the eccentric or inclined EMRIs orbits~\cite{Barsanti:2022ana,DellaRocca:2024pnm}.

From the asymptotic behaviour of the scalar field stress-energy tensor,
the energy and angular momentum fluxes can be read off \cite{Gralla:2005et,Warburton:2010eq}

\begin{equation}\label{fluxes:scal:eqs}
\begin{aligned}
\dot{E}_S^{\infty} &\equiv\Big(\frac{dE}{dt}\Big)^{\infty}_{\rm Scalar}  = \frac{q_s^2}{16\pi} \sum_{\ell mkn} \omega^2_{mkn} |Z^{s,+}_{\ell mkn}|^2, \\
\dot{E}_S^H&\equiv\Big(\frac{dE}{dt}\Big)^{H}_{\rm Scalar} = \frac{q_s^2}{16\pi} \sum_{\ell mkn} \omega_{mkn}\mathcal{K}_\omega |Z^{s,-}_{\ell mkn}|^2, \\
\dot{L}_S^{\infty} &\equiv\Big(\frac{dL}{dt}\Big)^{\infty}_{\rm Scalar}  = \frac{q_s^2}{16\pi} \sum_{\ell mkn} m \omega_{mkn} |Z^{s,+}_{\ell mkn}|^2, \\
\dot{L}_S^H&\equiv\Big(\frac{dL}{dt}\Big)^{H}_{\rm Scalar} = \frac{q_s^2}{16\pi} \sum_{\ell mkn} m \mathcal{K}_\omega |Z^{s,-}_{\ell mkn}|^2\;. \\
\end{aligned}
\end{equation}
Besides the two above fluxes, there is the averaged rate of change of the Carter constant, named as the Carter flux, derived with the Killing tensor equation in Refs.~\cite{Drasco:2005is, Isoyama:2013yor}, which is compatible with our symbols.
\begin{equation}\label{fluxes:scal}
\begin{aligned}
\dot{Q}_S^H &\equiv\Big(\frac{dQ}{dt}\Big)^{H}_{\rm Scalar} =  \sum_{\ell mkn} |Z^{s,-}_{\ell mkn}|^2 \frac{q_s^2}{16\pi}
\Bigg[ \Omega_r k  
 \\& 
+ 2\left(<V_{tr}>\omega^2 - m \mathcal{K}_\omega  <V_{\phi r}>\right)\Bigg]
\;,\\
\dot{Q}_S^\infty&\equiv\Big(\frac{dQ}{dt}\Big)^{\infty}_{\rm Scalar} = \sum_{\ell mkn}  2\left(\dot{E}_S^\infty <V_{tr}> - \frac{m}{\omega} \dot{E}_S^\infty <V_{\phi r}> \right) 
\\& +
\frac{q_s^2\Omega_r k}{16\pi}  |Z^{s,+}_{\ell mkn}|^2 
\;.\\
\end{aligned}
\end{equation}
Here, the symbol $<>$ denotes the average value over Mino time at one period, such as $<V_{\phi r}> = 1/\Lambda_r \int_0^{\Lambda_r}V_{\phi r}[r(\lambda)] d\lambda $ with a radial period $\Lambda_r = 2\pi/\Upsilon_r$.

We emphasize further again that the gravitational and scalar fluxes are computed within the standard Teukolsky formalism using Eqs.~\eqref{fluxes:grav:eqs}, \eqref{fluxes:scal:eqs} and \eqref{fluxes:scal}. We first solve the homogeneous radial Teukolsky equation with the MST method to obtain the ingoing and outgoing solutions that satisfy the physical boundary conditions at the horizon and at infinity. These homogeneous solutions are then used to evaluate the source integrals using the Green's function method, yielding the mode amplitudes $Z^{H,\infty}_{\ell mkn}$ (or equivalently $Z^{\pm}_{\ell mkn}$) for the gravitational and scalar perturbations. The energy, angular momentum and Carter fluxes are obtained from the asymptotic behavior of these amplitudes and summed over modes. All fluxes are computed numerically in \texttt{Mathematica}, and the resulting relativistic flux data are used to construct interpolation functions over the orbital parameter space for the subsequent adiabatic evolution. We introduce the interpolation scheme in the following subsections.

\subsection{Adiabatic evolution}\label{sec:4C}
For two subsections in Sec.~(\ref{grav:flux}) and (\ref{scal:flux}), we have briefly introduced the recipe of gravitational and scalar fluxes for eccentric and inclined EMRI orbits.
In this subsection, we will work within the adiabatic approximation to compute the inspiraling orbits, where the inspiral evolves more slowly than the orbital timescale due to the presence of gravitational and scalar radiations and
a tiny mass-ratio. This enables us to approximately regard the trajectory as
a sequence of geodesics, in which the energy, angular momentum and Carter constant losses are directly tied to the corresponding fluxes: 
\begin{equation}\label{fluxes:balance}
\begin{aligned}  
\dot{\mathcal{C}}^{\textup{(orbit)}}& \equiv ({d\mathcal{C}/dt})^{\textup{(orbit)}}
 = -\dot{\mathcal{C}}_S - \dot{\mathcal{C}}_G; \hspace{2mm}\mathcal{C} \in (E, L, Q)\;, \\
\dot{\mathcal{C}}_S  &= \dot{\mathcal{C}}_S^\infty + \dot{\mathcal{C}}^H_S; \hspace{2mm} 
\dot{\mathcal{C}}_G  = \dot{\mathcal{C}}_G^\infty + \dot{\mathcal{C}}^H_G \;.
\end{aligned}
\end{equation}
 Then we can use the gravitational and scalar fluxes in Eqs.~\eqref{fluxes:grav:eqs} and \eqref{fluxes:scal:eqs} to conduct the adiabatic evolution of the secondary object in EMRI, the orbital geometric parameters are derived by the following equations~\cite{Hughes:2021exa} 
\begin{equation}
\begin{aligned}\label{evolve:odes}
\frac{dp}{dt} &= \frac{(1-e^2)}{2}\frac{dr_a}{dt} +\frac{(1+e^2)}{2}\frac{dr_p}{dt}\,,\\
\frac{de}{dt} &= \frac{(1-e^2)}{2p} \Bigg[(1-e) \frac{dr_a}{dt} +(1+e)\frac{dr_p}{dt} \Bigg]\;,\\
\frac{dx}{dt} &= \mathcal{D}^{-1}\Big(2x(1-x^2)L (dL/dt)^{(\rm orbit)}- \\
& \hspace{2mm}x^3 [(dQ/dt)^{(\rm orbit)} +2(1-x^2)a^2E  (dE/dt)^{(\rm orbit)}]\Big),
\end{aligned}
\end{equation}
where $\mathcal{D}=2[L^2+a^2(1-E^2)x^4]$ and the explicit expressions of $\frac{dr_{a,p}}{dt}$ can be determined with the fluxes: $\frac{dr_{a,p}}{dt} = \Big(\frac{\partial E}{\partial r_{a,p}}\Big)^{-1}  \left(\frac{dE}{dt}\right)^{\textup{(orbit)}} + \Big(\frac{\partial L}{\partial r_{a,p}}\Big)^{-1} \left(\frac{dL}{dt}\right)^{\textup{(orbit)}} 
+ \Big(\frac{\partial Q}{\partial r_{a,p}}\Big)^{-1} \left(\frac{dQ}{dt}\right)^{\textup{(orbit)}}$.
When the orbital parameters $[p(t),e(t), x(t)]$ evolve with three differential equations in Eq.~\eqref{evolve:odes}, the three accumulated orbital phases can also be obtained as a result, 
\begin{eqnarray}
\Phi_{r,\theta,\phi} (t_{\rm obs}) = \int_{t_0}^{t_{\rm obs}} \Omega_{r,\theta,\phi}\Big[a,p(t),e(t),x(t)\Big] dt\;,
\end{eqnarray}
where $t_{\rm obs}$ is the time observed by the detectors during its operation and $t_0$ is the initial time, which is 
generally set to zero. Along with the definition of accumulated phases, the dephasings can be obtained with
\begin{equation}
\delta\Psi_{r,\theta,\phi} = \vert \Phi_{r,\theta,\phi}^{q_s=0} - \Phi_{r,\theta,\phi}^{q_s\neq0}\vert ,
\end{equation}
which correspond to the radial, polar and azimuthal dephasings. The quantities with the superscripts $(q_s=0, q_s\neq0)$ denote the dephasings computed with the standard GR
fluxes and the scalar charge modifying fluxes. According to the discussion on the distinguishability of dephasing \cite{Gupta:2021cno,Barsanti:2021ydd}, the threshold of dephasing recognized by LISA-like detectors is $\delta\Psi\sim0.1~\rm rad$ for the GW signal with a SNR $\sim30$~\cite{Bonga:2019ycj}.

To study the effect of the scalar charge on the EMRI dynamics, we adopt the flux-balance law, as mentioned above, to compute the adiabatic inspiraling orbits. Because the computation of the fully relativistic gravitational and scalar fluxes step by step is rather time-consuming and expensive, we adopt the flux-grid based scheme to generate gravitational and scalar fluxes fast, commonly used in EMRI studies \cite{Datta:2024vll, Kumar:2025njz}, in which the orbital evolution is driven by interpolating precomputed fluxes over a discretized parameter space. In particular, we first construct a three-dimensional grid for fixed values of the scalar charge and the spin of MBH, and compute the energy, angular momentum and Carter constant fluxes.
The grid totally contains $16\times16\times11$ points in three directions, sampling 16 points in the square of eccentricity $e^2\in[0, 0.35]$ and 16 points in the function $u(p)$ depending on the semi-latus rectum $p$, defined as: $u(p) = 1/\sqrt{p-0.9p_{\rm LSO}}$ with the $p_{\rm LSO}$ being the last stable orbit (LSO) in Kerr spacetime.
Together with this function $u$ in Ref.~\cite{Hughes:2021exa}, one can uniformly sample 16 points with the parameter $p$ between the maximum $p_{\rm max} = p_{\textup{LSO}}+12$ and the minimum 
$p_{\textup{min}}=p_{\textup{LSO}}+0.02$. For the inclination $x$ sampling, we take an arithmetic progression with 11 points $x\in[0.1,0.9]$. This grid configuration ensures a denser grid near the inner edge of the LSO, improving the robustness of interpolation on three-dimensional grids.
Since the fluxes change rapidly near the separatrix of LSO, we offset the grid slightly to maintain the validity of the adiabatic approximation. 
The small exclusion parameter region in the range of $p_{\textup{sp}}\leq p \leq p_{\textup{sp}}+0.02$ easily incurs some numerical errors, allowing stable and accurate numerical interpolation fluxes to run three-dimensional grids where the adiabatic method still holds.
Note that including higher modes significantly increases the computational cost, particularly when using a three-dimensional inspiraling grid, as discussed in detail in Ref.~\cite{Kumar:2025njz}. 
Therefore, we focus on the summation of fluxes by setting the truncation maximum index to $(\ell, m, k, n)_{\text{max}} = (2, 2, 2, 2)$ for the computation of both the gravitational and scalar fluxes. 
For the scalar flux, the truncation maximum index is also setting to  $(\ell, m, k, n)_{\text{max}} = (2, 2, 2, 2)$,
this truncation includes the dominant $(\ell, m) = (1, 1)$ mode. These modes capture the leading contributions from gravitational quadrupole and scalar dipole radiation to the orbital dynamics in a wide range of binary black hole configurations relevant to our analysis~\cite{OBrien:2019hcj}. 
The truncation was chosen to balance physical completeness with computational feasibility in a three-dimensional inspiral grid over ($p, e, x$). Additionally, as our analysis focuses on generic eccentric and inclined orbits, for which eccentricity activates higher $n$-modes, while inclination activates nonzero $k$-modes. As a result, the scalar energy flux is distributed across multiple ($\ell, m, k, n$) harmonics rather than being dominated by a single dipole mode, particularly in the strong-field, periapsis-dominated portion of the inspiral and when the dynamics is exhibiting generic orbits. Since, extending the mode sum beyond $(\ell, m, n, k)_{\mathrm{max}}$ = ($2, 2, 2, 2$) substantially increases the computational cost of constructing three-dimensional flux grids, which remains a focus of future investigation with computationally more efficient methods, we therefore set the modes up to $(2, 2, 2, 2)_{\mathrm{max}}$.


\subsection{Spline interpolation and error control}
With the EMRI flux data obtained via the perturbation theory, we compute the gravitational and scalar fluxes at any point within the aforementioned sampled grid range  using an interpolation scheme. The \texttt{GNU Scientific Library (GSL)}, implemented in \texttt{C++}, is employed to efficiently generate these fluxes within the package of \texttt{FastEMRIWaveforms (FEW)}~\cite{Katz:2021yft, Chua:2020stf}.
To efficiently evaluate the radiation fluxes required for adiabatic EMRI evolution, we construct a three-dimensional spline interpolant for fluxes
\begin{equation}
    \mathcal{F}\bigl(u(p), e^{2}, x\bigr),
\end{equation}
where \(u(p)\) is the transformed semi-latus rectum coordinate, \(e^{2}\) denotes the squared eccentricity, and \(x=\cos I\) is the inclination parameter defined in Sec.~\ref {sec:4C}.  
The exact fluxes are computed on a \(16\times 16\times 11\) grid covering the full domain of interest.  
To stabilize the relative accuracy across several orders of magnitude, the interpolation is performed on \(\log_{10}\!\left|\mathcal{F}\right|\).  
For each fixed value of \(x\), a bicubic spline (implemented using \texttt{gsl\_spline2d\_bicubic}) is constructed in the \((u,e^{2})\) plane, and the resulting values are subsequently interpolated along \(x\) using a one-dimensional cubic spline (implemented with \texttt{gsl\_spline\_cspline}).
To assess the  global interpolation accuracy, we compute the exact fluxes
\(\mathcal{F}_{\mathrm{true}}\) and the spline-interpolated fluxes \(\mathcal{F}_{\mathrm{interp}}\) at a large set of randomly sampled and off grid-points spanning the entire parameter domain, and we monitor the relative deviation
\begin{equation}
    \epsilon_{\mathrm{rel}}
    = \left|
        \frac{\mathcal{F}_{\mathrm{true}}
              -\mathcal{F}_{\mathrm{interp}}}
             {\mathcal{F}_{\mathrm{true}}}
      \right|,
\end{equation}
where $\mathcal{F}\in\{\dot{E}_{S,G}^{\infty,H},\dot{L}_{S,G}^{\infty,H},\dot{Q}_{S,G}^{\infty,H} \}$ is used to denote the symbol of the scalar or gravitational flux at the infinity and near the horizon.
For the adopted grid, the interpolation error satisfies
\begin{equation}
\mathcal{F}_{\mathrm{rel}}^{\rm error}
    = \max \epsilon_{\mathrm{rel}}
    \;\lesssim\; 10^{-6},
\end{equation}
with typical errors lying in the range \(10^{-7}\!-\!10^{-6}\).  
If the accuracy criterion is not satisfied at specific grid points or at off-grid validation locations, we refine the sampling locally by computing additional perturbative fluxes in the vicinity of those points. The spline interpolant $\mathcal{F}_{\rm interp}$ is then rebuilt following the procedure described above, and the validation step is repeated until the required tolerance is achieved. The resulting pre-validated flux table provides a smooth and numerically stable representation of $\mathcal{F}(u,e^{2},x)$, and is well suited for long-duration inspiral evolution as well as high-precision parameter-estimation studies.

\begin{table}[htbp!]
\centering
\begin{tabular}{cc|ccc}
\hline
\hline
$q$ & $x$ & $\mathcal{F}_{\mathrm{rel}}^{\rm error}$ & $\delta \Psi_\phi [\rm rad] $ &$\mathcal{M}$   
\\
\hline
$10^{-4}$  & 0.1
&$1.4\text{e-6}$    &$0.78$    &$1.29\text{e-2}$
\\
&0.3 
&$3.4\text{e-6}$  &$1.87$      &$3.45\text{e-2}$   
\\
&$0.5$   
&$4.7\text{e-6}$  &$3.547$      &$4.78\text{e-2}$   
\\
\hline
\hline
$10^{-5}$  & 0.1
&$1.4\text{e-6}$    &$0.152$    &$2.67\text{e-4}$
\\
&0.3 
&$3.4\text{e-6}$  &$0.367$      &$4.35\text{e-4}$   
\\
&$0.5$   
&$4.7\text{e-6}$  &$0.547$      &$5.54\text{e-4}$   
\\
\hline
\hline
$10^{-6}$  & 0.1
&$1.4\text{e-6}$    &$0.0145$    &$2.45\text{e-5}$
\\
&0.3 
&$3.4\text{e-6}$  &$0.0784$      &$4.75\text{e-5}$   
\\
&$0.5$   
&$4.7\text{e-6}$  &$0.0678$      &$5.32\text{e-5}$   
\\
\hline
\hline
\end{tabular}
\caption{The effect of our adopted interpolants on the observable physical quantities is listed quantitatively. We here consider the inspiral of binaries with $M=10^6M_\odot$, $\mu=10M_\odot$, $a=0.9$, $p_0=12$ and $e_0=0.3$, and comparisons of the adiabatic evolutions between the interpolations and 5PN analytic fluxes is listed \cite{Isoyama:2021jjd}, which are the azimuthal dephasing and mismatch using two method.
}\label{tab:interpolation:mis}
\end{table}

In order to investigate the effect of interpolation-flux on the accuracy of an adiabatic inspiral, we report the azimuthal dephasing and mismatch against the GSL-based interpolation in the Table~\ref{tab:interpolation:mis}. To simplify the computational resources, we only consider EMRIs driven by the gravitational radiation, which also serve as an indicator of the robustness of our GSL-based interpolant. According to the results in above Table ~\ref{tab:interpolation:mis}, 
we can find that, for a typical EMRI with mass-ratio $q=10^{-5}$ and inclined orbits,  the azimuthal dephasing is less than one radian and the mismatch is $<10^{-3}$. The case of mass-ratio $q=10^{-4}$ would generate a bigger deviation that dephasing is $>1$ radian and mismatch is $>0.00125$, while the case of mass-ratio $q=10^{-6}$ ensures a better accuracy of inspirals.
Therefore, we should generate the adiabatic inspirals using the interpolated-flux carefully for different mass-ratio EMRI systems.

\section{Waveform and Mismatch}\label{waveform&mismatch}
When the orbital parameters of the inspiraling secondary are evolved with the gravitational and scalar emission reaction, the GW waveform emitted from EMRIs can be efficiently generated within the \texttt{FEW} framework \cite{Katz:2021yft, Barack:2003fp, Chua:2020stf} to examine the effects of the scalar charge on the EMRI signal.
For a rotating massive BH, \texttt{FEW} can import the \texttt{Augmented Analytical Kludge (AAK)} module to compute the response of the LISA signal in the low-frequency regime \cite{Chua:2017ujo}. More recently, the latest version of the \texttt{FEW} model has extended the fully relativistic waveforms to Kerr MBHs, which can efficiently generate the EMRI signal to meet the requirements of GW data analysis~\cite{Chapman-Bird:2025xtd}. 
\texttt{AAK} module  generates the waveform, where the two polarization modes, plus and cross $(+,\times)$, are written as the summation of a series of harmonic components in the transverse-traceless gauge 
\begin{align}\label{amplitude}
h_+ \equiv \sum_n A_n^+ = \sum_n &-\Big[1+(\hat{L}\cdot\hat{n})^2\Big]\Big[a_n \cos2\gamma -b_n\sin2\gamma\Big] \nonumber\\
& +c_n\Big[1-(\hat{L}\cdot\hat{n})^2\Big], \\
h_\times \equiv \sum_n A_n^\times = \sum_n &2(\hat{L}\cdot\hat{n})\Big[b_n\cos2\gamma+a_n\sin2\gamma\Big]\;,\nonumber
\end{align}
where ($\hat{n}, \hat{L}$) is the unit vector along the source direction and orbital angular momentum, respectively, and
their dot product can be given by four location angles  $(\theta_L, \phi_L,\theta_K, \phi_K)$
\begin{eqnarray}
\hat{L}\cdot\hat{n} = \cos\theta_S\cos\theta_L + \sin\theta_S \sin\theta_L \cos(\phi_S-\phi_L).
\end{eqnarray}
More detailed descriptions of the angles $(\theta_L, \phi_L,\theta_K, \phi_K)$ can be found in \cite{Barack:2003fp}.
The coefficients ($a_{n}, b_{n}, c_{n}$) can be recasted into eccentricity-dependent functions as follows \cite{Peters:1963ux}
\begin{equation}
\begin{aligned}
a_n =~ &-n \mathcal{A} \Big[J_{n-2}(ne)-2eJ_{n-1}(ne)+\frac{2}{n}J_n(ne) \\
& +2J_{n+1}(ne) -J_{n+2}(ne)\Big]\cos(n\Phi_r), \\
b_n =~ &-n \mathcal{A}(1-e^2)^{1/2}\Big[J_{n-2}(ne)-2J_{n}(ne)+J_{n+2}(ne)\Big]\\
& \times\sin(n\Phi_r), \\
c_n =~& 2\mathcal{A}J_n(ne)\cos(n\Phi_r) \hspace{1mm} ; \hspace{1mm} \mathcal{A} = ~ (2\pi M \omega_\phi )^{2/3}m_p/d_L,
\end{aligned}
\end{equation}
where $J_n$ are the Bessel function of the first kind and $\gamma = \Phi_\phi - \Phi_r$ represents the direction of eccentric orbit pericenter ~\cite{Barack:2003fp}, $d$ is the distance from the source to the detector.
Thus, the low-frequency EMRI signal detected by the space-based detector LISA, TianQin and Taiji, can be approximately given with the antenna pattern functions ($F^{+,\times}_{I,II}$) \cite{Cutler:1994ys} in the following way
\begin{equation}
h_{\rm I,II} = \frac{\sqrt{3}}{2} (F^+_{\rm I,II} h^+ + F^\times_{\rm I,II} h^{\times}).
\end{equation}
The complete details of the related expressions can also be found in \cite{PhysRevD.69.082005}. 
\begin{figure*}[htb!]
\centering
\includegraphics[width=3.2in, height=2.2in]{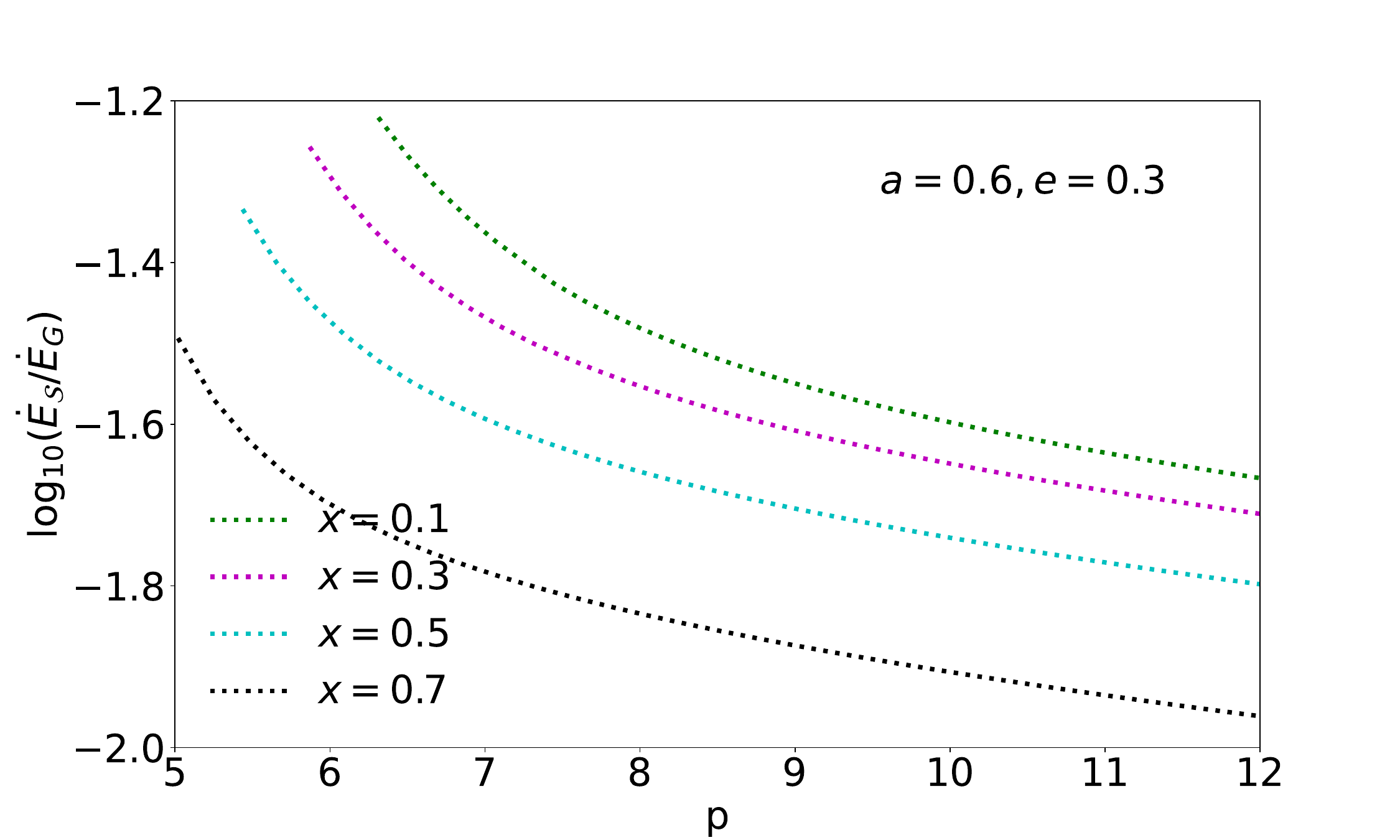}
\includegraphics[width=3.2in, height=2.2in]{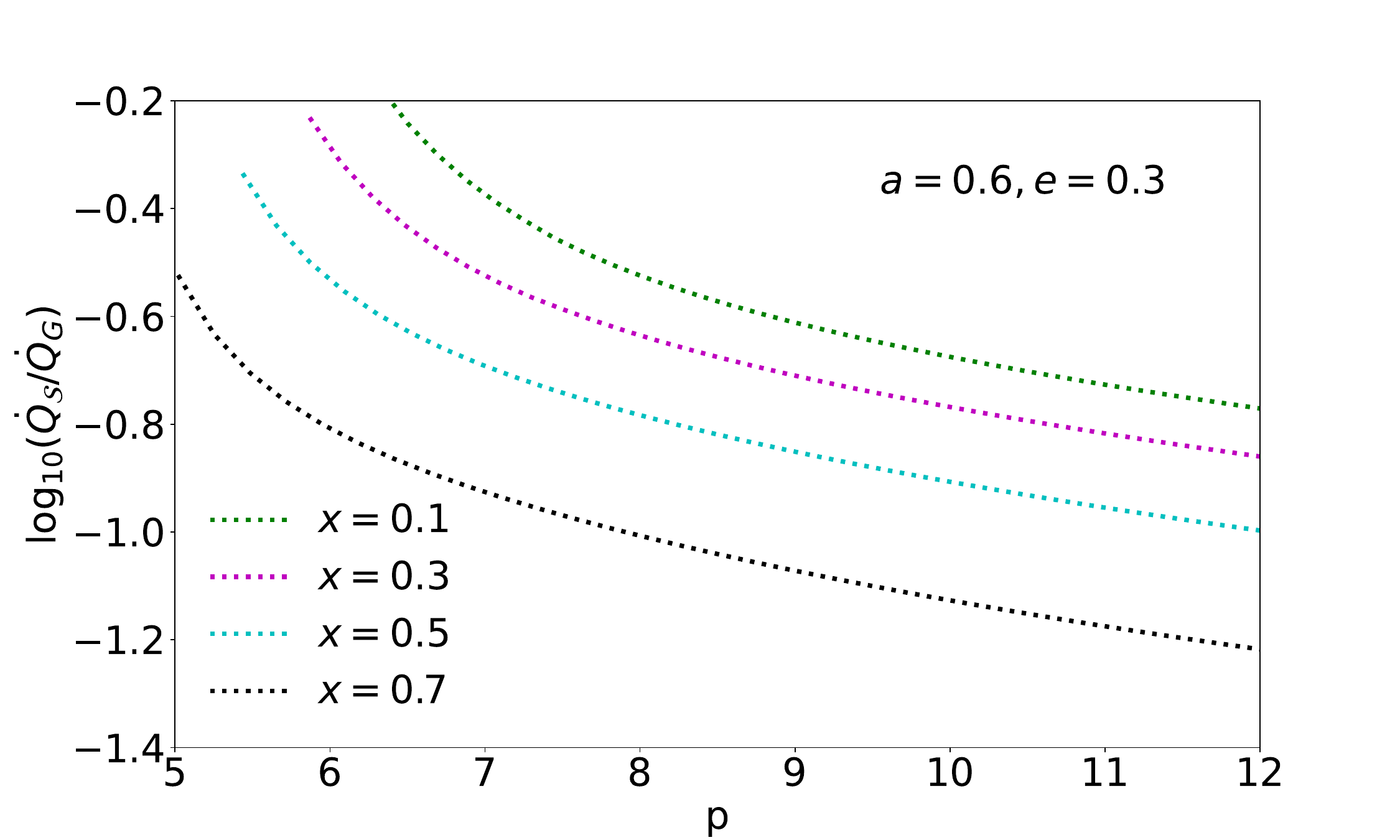}
\includegraphics[width=3.2in, height=2.2in]{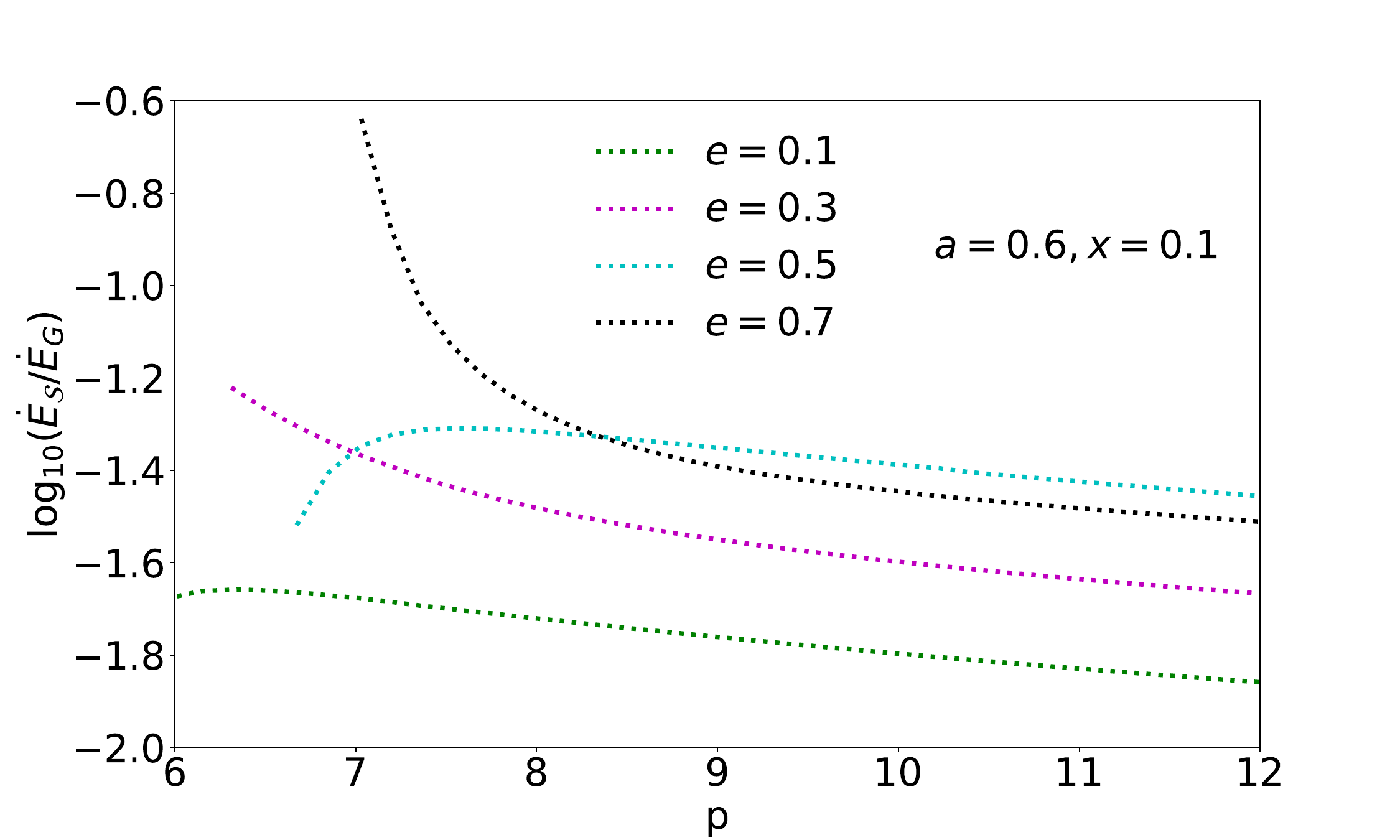}
\includegraphics[width=3.2in, height=2.2in]{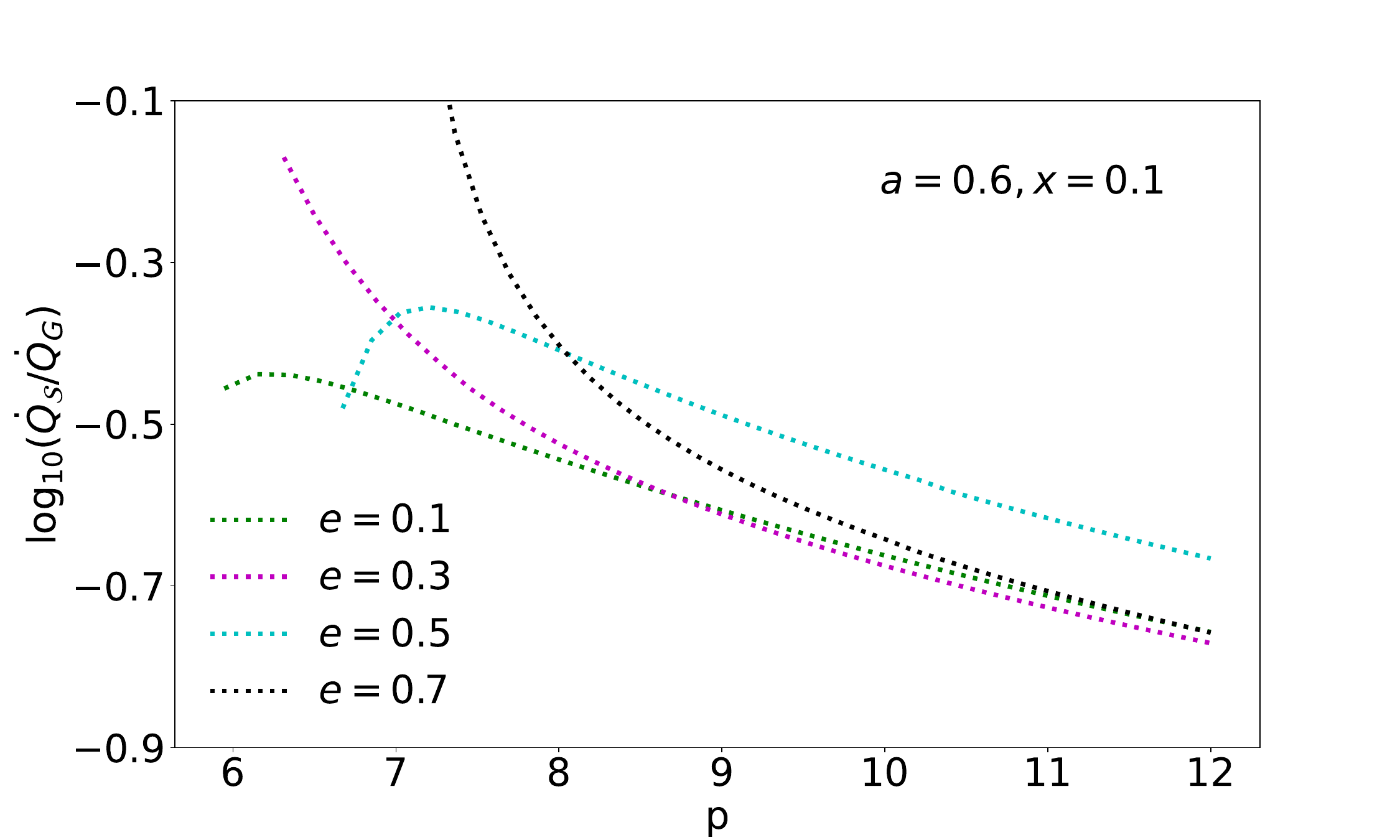}
\caption{Ratios between the gravitational and scalar fluxes,
relative ratio of changing rates of energy and Carter constant $\big[\log_{10}(\dot{E}_S/\dot{E}_G), \log_{10}(\dot{Q}_S/\dot{Q}_G)\big]$, are plotted for different orbital parameter settings and a fixed spinning MBH with $a=0.6$. The top panels show the fluxes ratio of the energy and Carter constant for a fixed eccentricity $e=0.3$, distinct orbital inclination $x\in[0.1,0.3,0.5,0.7]~(I\in[0.47,0.4,0.33,0.24]\pi)$. The bottom panels are also the fluxes ratio for a fixed orbital inclination $x=0.1$ and the changing eccentricity $e\in[0.1,0.3,0.5,0.7]$.
The other parameters are set as follows: the mass-ratio of the binary $q=10^{-5}$, the scalar charge $q_s=0.05$.} \label{fig:flux:ratio}
\end{figure*}


\begin{figure*}[htb!]
\centering
\includegraphics[width=3.4in, height=3.09in]{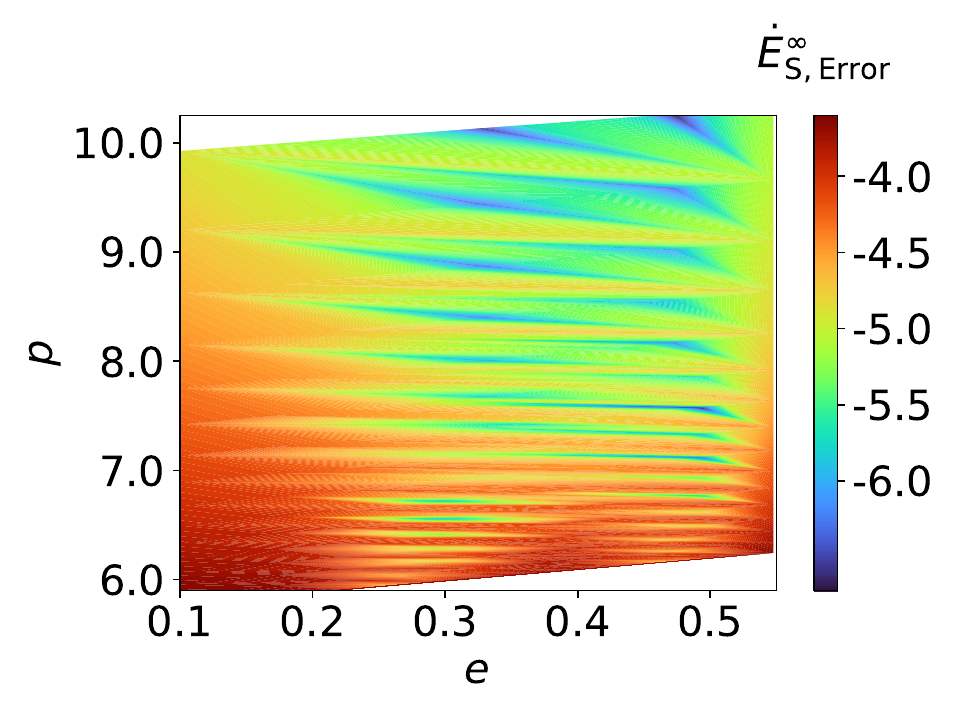}
\includegraphics[width=3.4in, height=3.09in]{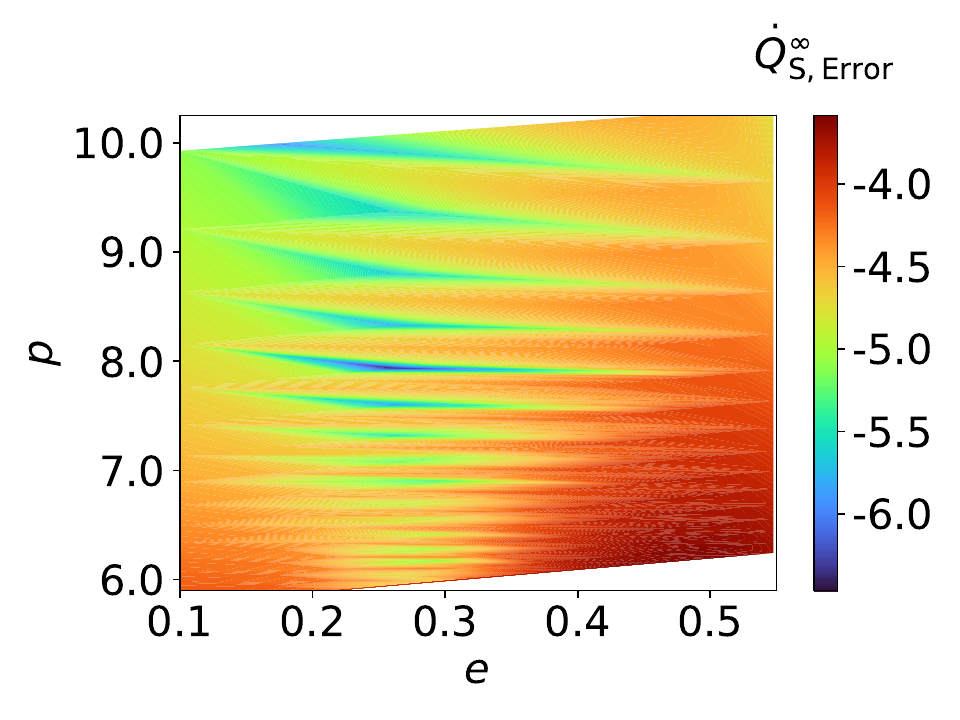}
\caption{Interpolation error of scalar energy and Carter fluxes as the function of orbital parameters $(p,e)$ for the rotating MBH with a dimensionless spin $a=0.3$ is plotted, setting orbital inclination $x=0.1$. These are the difference of fluxes from the relativistic and interpolation methods over the sampling grids. The other parameters are set as follows: the mass-ratio of the binary $q=10^{-5}$, the scalar charge $q_s=0.05$.} \label{fig:flux:error}
\end{figure*}

\begin{figure*}[htb!]
\centering
\includegraphics[width=3.85in, height=3.09in]{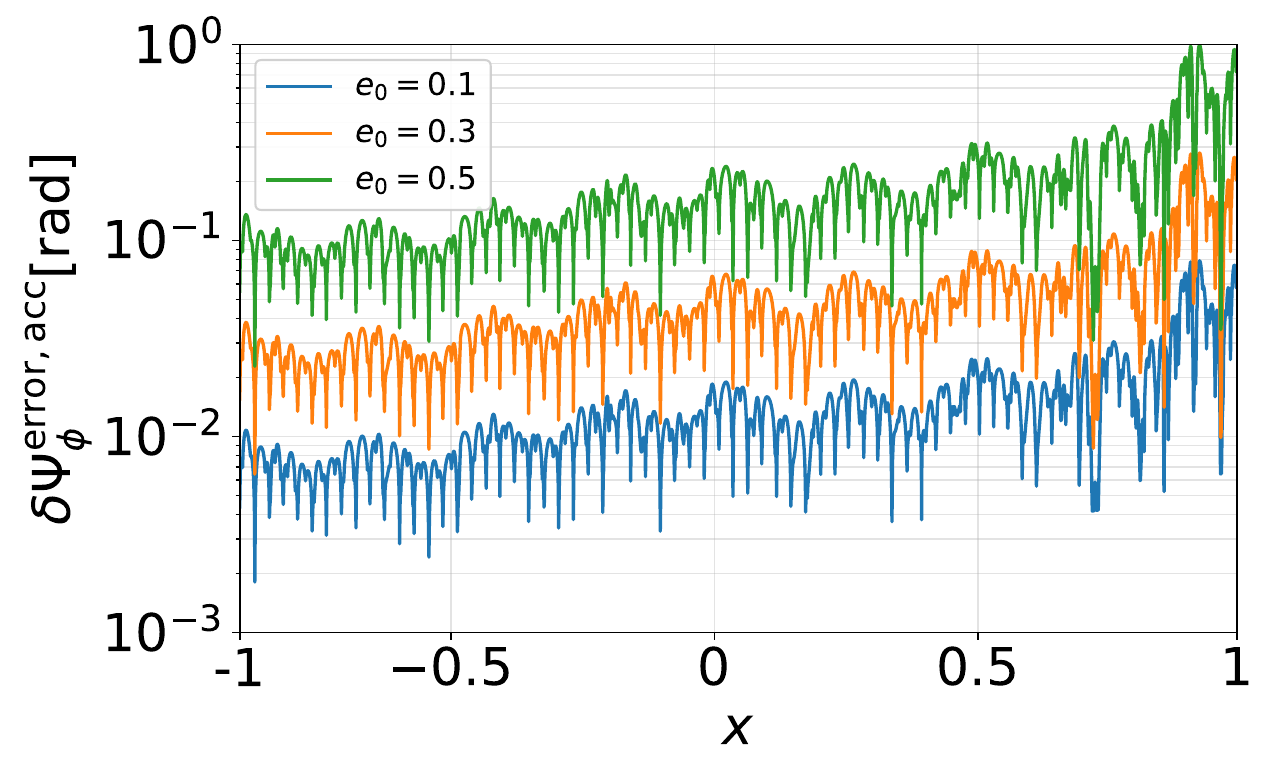}
\caption{Maximum value of accumulated phase error induced by the imprecision of interpolated gravitational energy flux as function of orbital inclination is plotted, including the inspiral of two-year and three cases of eccentricities $e_0\in\{0.1,0.3,0.5\}$.} \label{fig:local:flux:acc:error}
\end{figure*}
In order to explore further the effect of scalar charge on the EMRIs waveforms, we conduct a mismatch analysis between waveforms with and without the scalar charge parameter, also termed as the standard GR and beyond GR cases.
This approach enables us to quantify the extent to which the presence of the scalar charge alters the waveform morphology.
For two waveforms, $h_{a}$ (with correction for scalar charge $q_s$) and $h_{b}$ (GR), their mismatch can be determined with the overlap $\mathcal{O}$ \cite{Cutler:1994ys}
\begin{align}\label{mismatch}
\mathcal{M}(h_{a}, h_{b}) = 1 - \mathcal{O}(h_{a}, h_{b}),
\end{align}
where the overlap and inner product between two waveforms in the frequency domain are given by
\begin{equation}
\begin{aligned}\label{overlap}
\mathcal{O}(h_a, h_b) =& \frac{(h_{a}\vert h_{b})}{\sqrt{(h_{a}\vert h_{a})(h_{b}\vert h_{b})}}, \\
(h_a|h_b) =& 2 \int^{f_{\rm high}}_{f_{\rm low}} \frac{h_a^*(f)h_b(f)+h_a(f)h_b^*(f)}{S_n(f)}df.
\end{aligned}
\end{equation}
where $f_{\textup{low}}=0.1~\textup{mHz}$ and $f_{\textup{high}}$ is the orbital frequency near the LSO. $S_{n}(f)$ is the power spectral density of LISA \cite{LISA:2017pwj} as shown in Appendix (\ref{appC}). Note that in our analysis, the LISA noise curve is coming from the LISA Pathfinder, alternatively, we adopt the standard LISA instrumental sensitivity curve derived from the LISA mission design specifications \cite{LISA:2017pwj, Babak:2017tow, Robson:2018ifk}, without including the white-dwarf confusion background \cite{Robson:2018ifk, Maselli:2021men}. The mismatch exhibits the degree of deviation between two signals and thus serves as an indicator to assess the distinguishability of the scalar charge effect within the sensitivity range of the LISA-like detectors. 
The mismatch is computed in the frequency domain, where the inner product is weighted by the power spectral density \( S_n(f) \) of LISA-like detectors. The frequency-domain waveform \( \tilde{h}(f) \) is obtained from the direct Fourier transform of the time-domain signal, where the inspiraling trajectory that is obtained with the Runge-Kutta method is injected into the quadrupole waveform formula in FEW model. The Fourier transform from the time-domain to frequency-domain signal is using the function of $\texttt{numpy.fft.fft}$ in the community of
$\texttt{python}$, without applying the window functions.

In this work, we employ the mismatch to evaluate the influence of a scalar charge on EMRI waveforms and the resulting implications for potential inferences. Our analysis does not incorporate the detailed time-delay interferometry (TDI) response, focusing instead on the intrinsic waveform differences induced by the scalar charge. Consequently, the detector power spectral density \( S_n(f) \) explicitly enters the computation of the inner product used in the mismatch evaluation. Note that the mismatch for generic EMRI orbits beyond GR framework specifically comes from two perspectives: one by implementing the standard Teukolsky equation of GR and the second with the modification of scalar charge carried by secondaries. According to Eq. (\ref{mismatch}), a mismatch $\mathcal{M}=0$ means two identical waveforms with maximum overlap $\mathcal{O}=1$. In true settings, a smaller deviation, such as the scalar charge in astrophysical environments~\cite{Dyson:2025dlj,Li:2025ffh}, can alter the waveform nature from the GR case. To distinguish the effect of scalar charge, a threshold $\mathcal{M}\leq 1/(2\rho^{2})$ is used \cite{Flanagan:1997kp,Lindblom:2008cm}, where $\rho$ is the signal-to-noise ratio (SNR). For $\rho=20$, this gives $\mathcal{M}_c\approx 0.00125$ \cite{Babak:2017tow, Fan:2020zhy}, also delimiting a benchmark for identifying whether the potential deviations from GR, scalar charge in the present context, are observable from next-generation detectors, serving as as a probe for determining signatures of modified gravity theories.

In GW parameter estimation, the FIM provides a standard and computationally inexpensive approximation to the expected measurement uncertainties of source parameters. Unlike Bayesian approaches based on Markov chain Monte Carlo (MCMC) sampling, the FIM does not require intensive exploration of high-dimensional priors and a larger amount of computational resources, it is therefore well suited for rapid forecasting studies. The FIM $\mathbf{\Gamma}$ is defined through the noise-weighted inner product as
\begin{equation}
\Gamma_{ij} \equiv 
\left( \frac{\partial h}{\partial \theta_i} \,\bigg|\, 
       \frac{\partial h}{\partial \theta_j} \right),
\qquad i,j = 1, \dots, 15 \, ,
\end{equation}
where $(\cdot|\cdot)$ denotes the usual detector inner product weighted by the one-sided noise power spectral density,
\begin{align}\label{ovp2}
(h_{1}|h_{2}) = 4\,\mathrm{Re}\!\left[\int_{0}^{\infty} 
\frac{\tilde{h}_{1}(f)\,\tilde{h}^{*}_{2}(f)}{S_{n}(f)}\,df \right],
\end{align}
with $\tilde h_i(f)$ being the Fourier transform of $h_i(t)$ and $S_n(f)$ the one-sided noise spectral density and all source parameters can be defined as
\begin{equation}
\theta_i =\{M,\mu,a,p,e,x,q_s,\Phi_\phi,\Phi_r,\Phi_\theta,\phi_S,\theta_S, \Phi_K,\theta_K, d\}\;.
\end{equation}
Under the high-SNR and locally Gaussian-likelihood approximation, the covariance matrix is given by the inverse Fisher matrix,
\begin{equation}\label{eq:cov:matrix}
\Sigma_{ij} \equiv 
\langle \delta\theta_i \, \delta\theta_j \rangle
= \left( \Gamma^{-1} \right)_{ij},
\end{equation}
and the corresponding $1\sigma$ uncertainty on each parameter is obtained from the diagonal entries,
\begin{equation}\label{sigma:fim}
\sigma_{\theta_i} = \sqrt{\Sigma_{ii}} \, .
\end{equation}
In this work, the FIM framework provides a first estimate of LISA's ability to constrain the scalar charge $q_s$. In addition, the off-diagonal elements of $\Sigma_{ij}$ quantify correlations among intrinsic and extrinsic parameters, thereby indicating potential degeneracies in the EMRI parameter space.

\begin{figure*}[htb!]
\centering
\includegraphics[width=3.3in, height=2.2in]{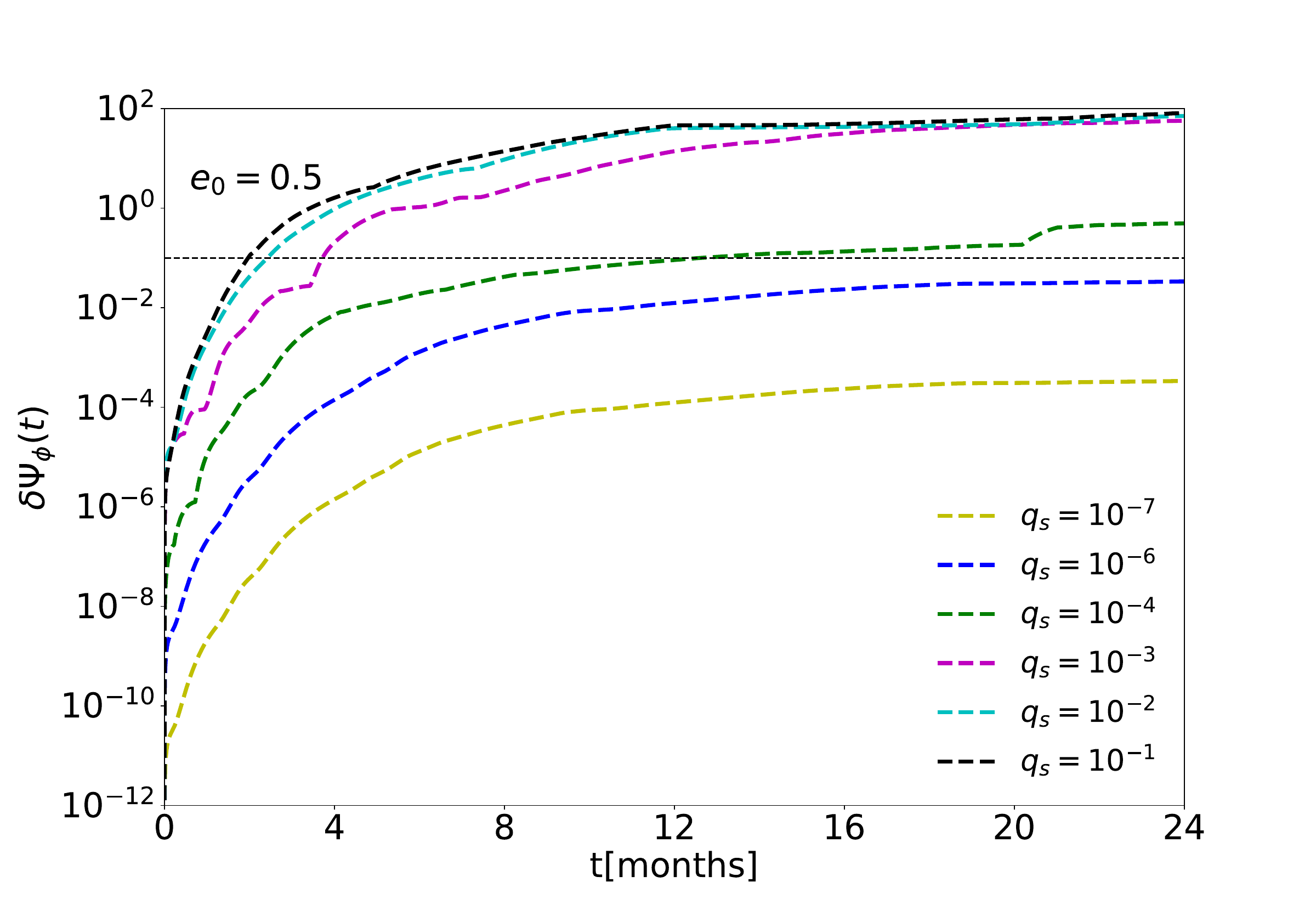}
\includegraphics[width=3.3in, height=2.2in]{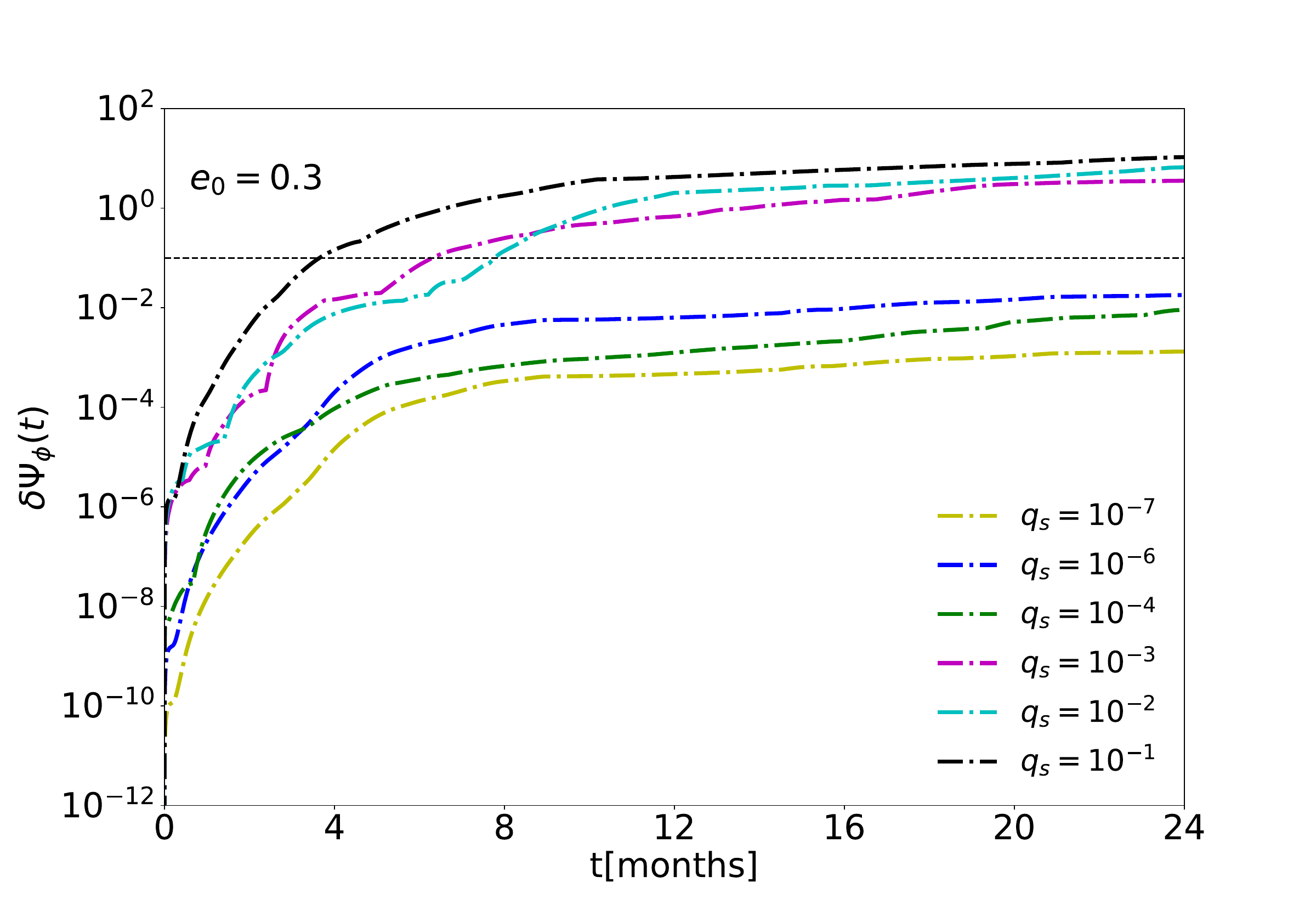}
\includegraphics[width=3.3in, height=2.2in]{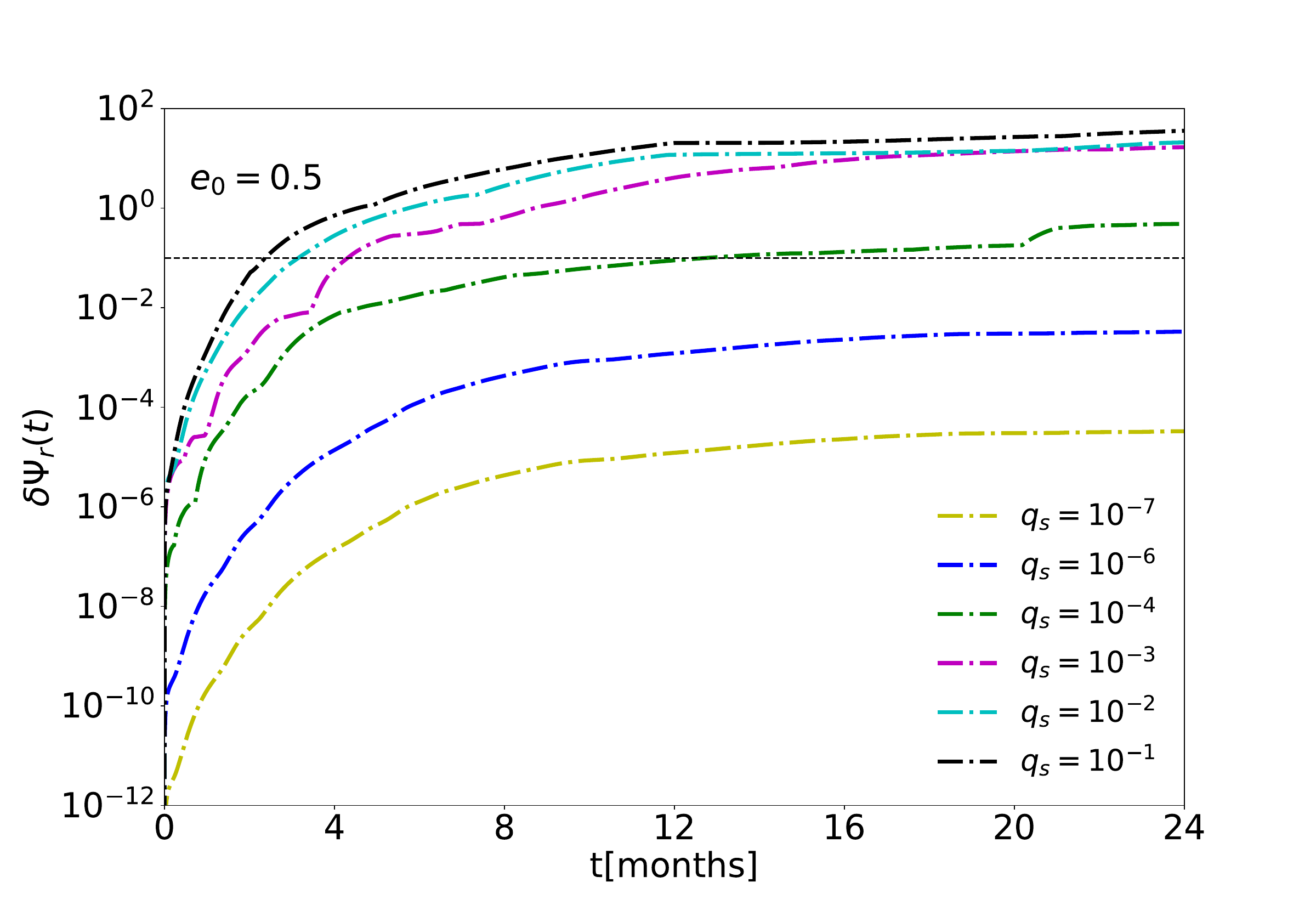}
\includegraphics[width=3.3in, height=2.2in]{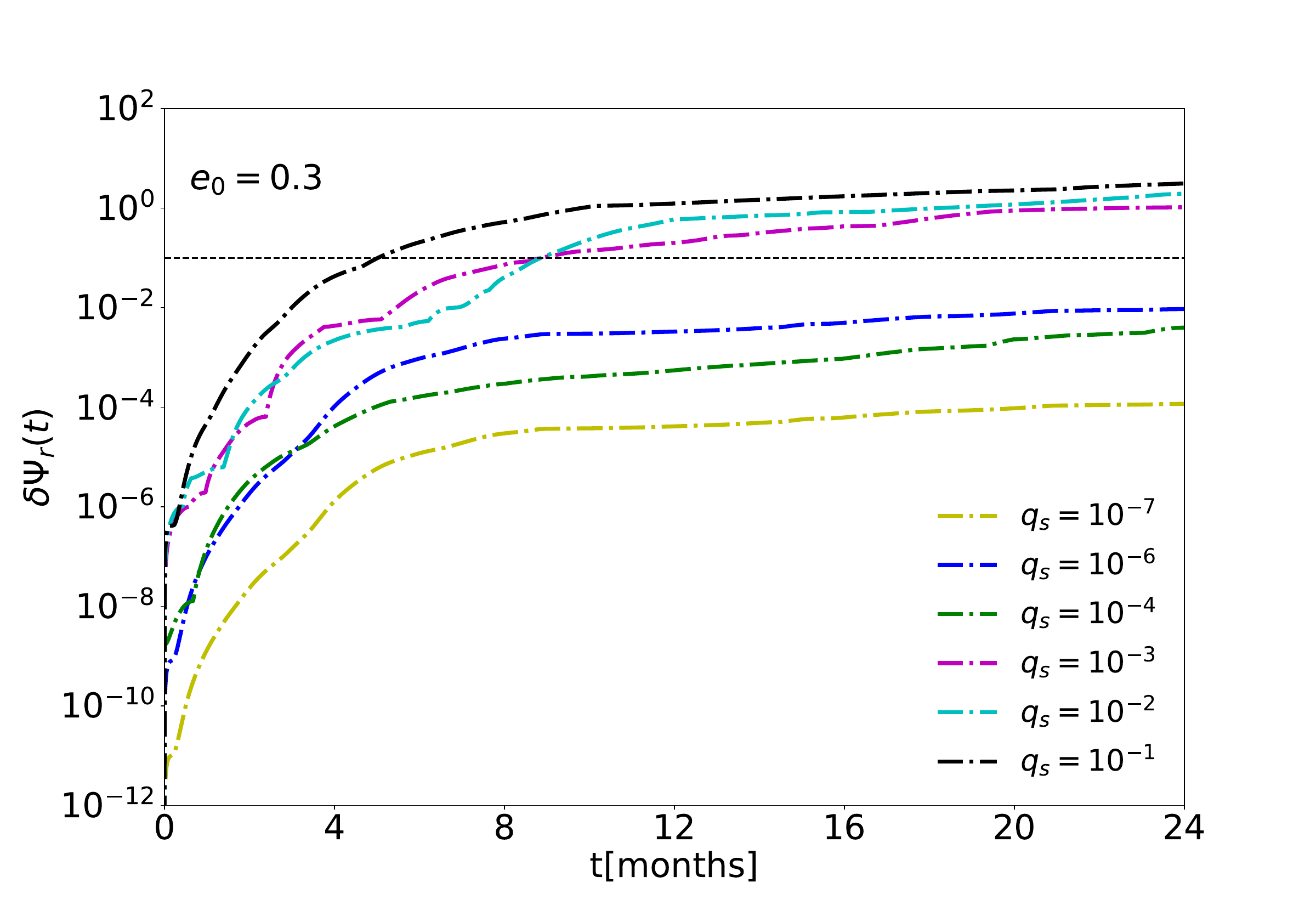}
\includegraphics[width=3.3in, height=2.2in]{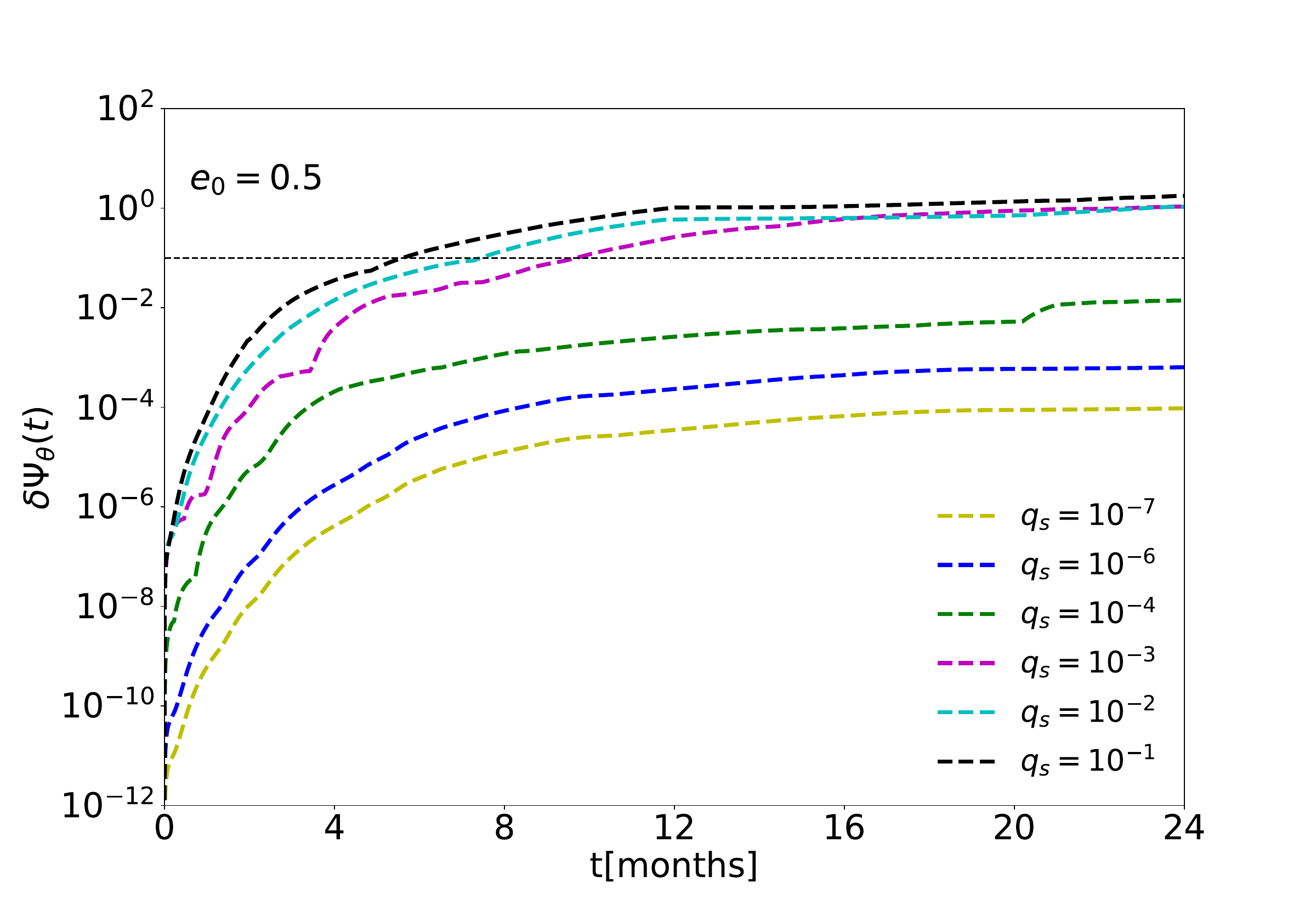}
\includegraphics[width=3.3in, height=2.2in]{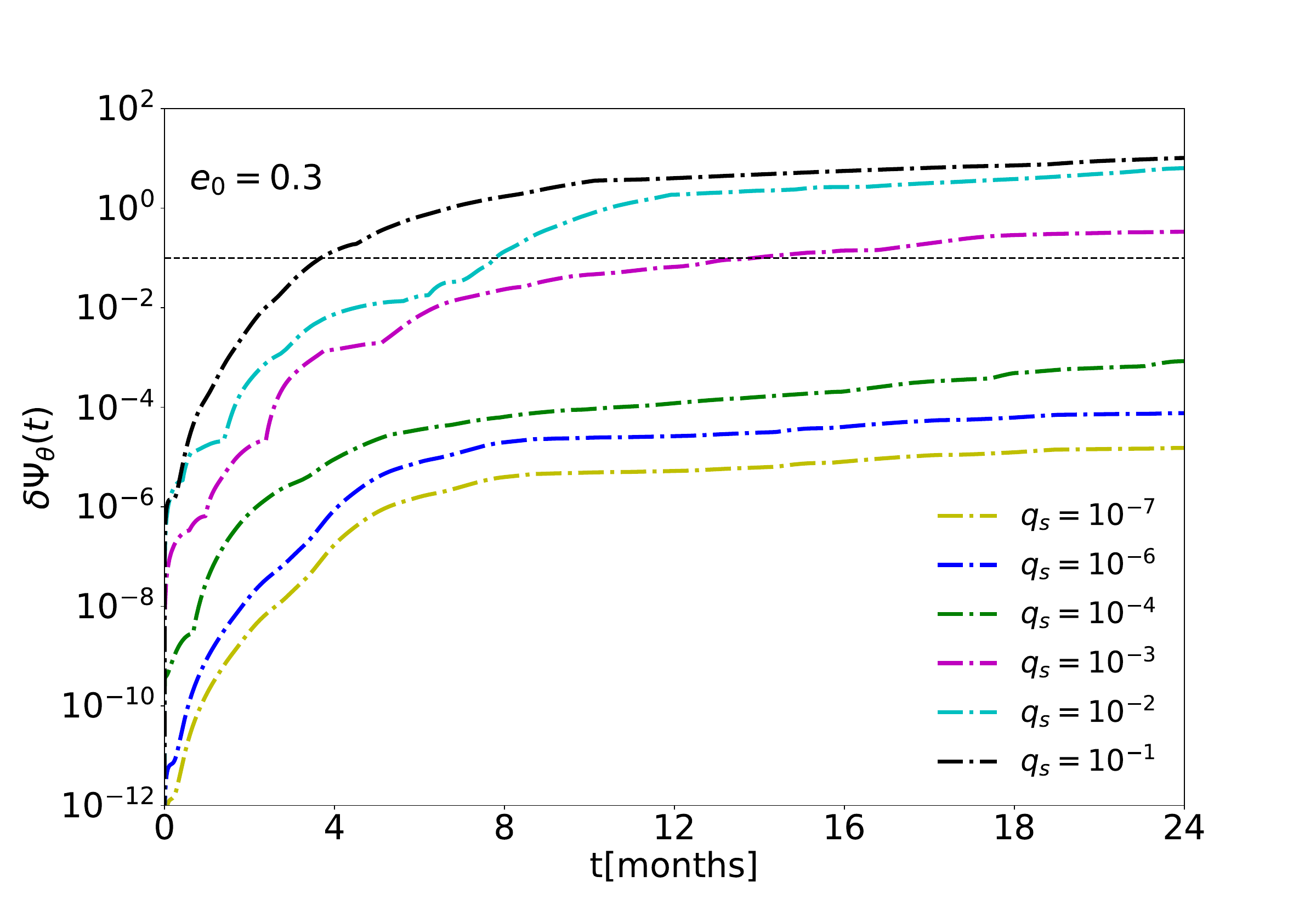}
\caption{Azimuthal, radial and polar dephasing as a function of observation time for two eccentricities $e=(0.3,0.5)$ and a fixed MBH spin $a=0.3$ is plotted, including six scalar charges $q_s=(10^{-7},10^{-6},10^{-4},10^{-3},10^{-2},10^{-1})$, where the observation time is set for two years. The horizontal black dashed line is the detection threshold distinguished LISA, which is about $0.1$ rad. The orbital initial eccentricity and inclination is $p=12$ and $x=0.5~ (I\sim0.33\pi)$, initial orbital phases are taken as follows $\Phi_{r,0} =\Phi_{\theta,0} =\Phi_{\phi,0}=1.0$
and the mass-ratio of binary system is $10^{-5}$.} \label{fig:dephasings}
\end{figure*}
\section{Results} \label{wave}
\subsection{Scalar fluxes and their effect on inspirls}
Here, we argue how the scalar charge modifies the orbital evolution of EMRI and the difference in waveform for two gravitational frameworks. In Fig.~\ref{fig:flux:ratio}, we place the ratio between the gravitational and scalar fluxes as a function of the semi-latus rectum with distinct values of orbital parameters, including the logarithm of ratio of the energy and Carter constant fluxes. The top pictures show two cases of $\log_{10}(\dot{E}_S/\dot{E}_G)$ for the left panel and $\log_{10}(\dot{Q}_S/\dot{Q}_G)$ for the right panel, setting four orbital inclinations $x=(0.1,0.3,0.5,0.7)$ and a fixed eccentricity $e=0.3$. We observe that the ratios of two fluxes increase as the orbital parameter $(I)$ become smaller, indicating the scalar fluxes play a vital role in the orbital evolution of inclined EMRIs.
On the other hand, the bottom figures show the cases for different eccentricities $e=(0.1,0.3,0.5,0.7)$ and a fixed orbital inclination $x=0.1~(I\sim 0.47\pi)$.
We notice that the ratios of the fluxes do not increase monotonously and the proportion of scalar fluxes grows until $e\gsim 0.5$. 
This behavior is different from the conclusion of the relation between the scalar and gravitational fluxes for the equatorial and eccentric orbits~\cite{Barsanti:2022ana},
it may be due to the fact that the inclusion of orbital inclination leads to a more complex situation and is naturally more relevant for the observations. A comparison/checks of fluxes with the existing literature is provided in Table (\ref{tab:fluxes:compare}) of the appendix (\ref{flx comp}).

When carrying out the evolution of EMRI orbits under the back-reaction of the gravitational and scalar field, we employ the interpolation method supplemented in \texttt{FEW} model to generate two fluxes more efficiently. So we assess the robustness of the interpolation method computing the fluxes over the three-dimensional grids in Fig.~\ref{fig:flux:error}.
Specifically, the errors of the fluxes from two methods (relativistic theory and interpolation method) as a function of orbital geometric parameters $(p,e,x)$ are plotted for a fixed MBH's spin $a=0.3$, including the interpolation error of energy (left panel) and Carter constant (right panel) fluxes. From two panels, the flux errors between two methods are distributed in the range of $[10^{-6},10^{-4}]$, allowing us to evolve orbital parameters derived by two kinds of interpolation fluxes. 
\begin{figure*}[htb!]
\centering
\includegraphics[width=3.0in, height=2.0in]{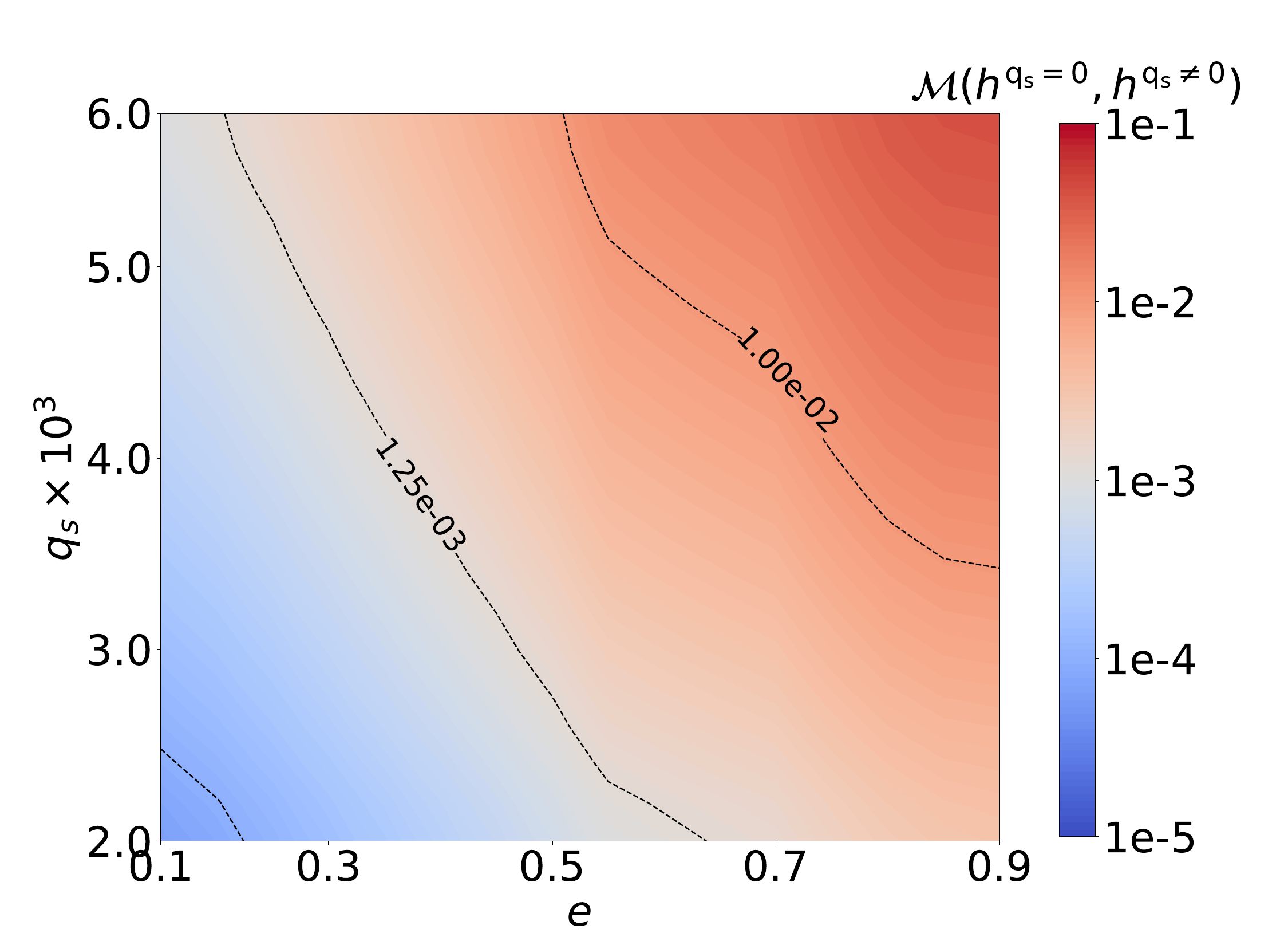}
\includegraphics[width=3.0in, height=2.0in]{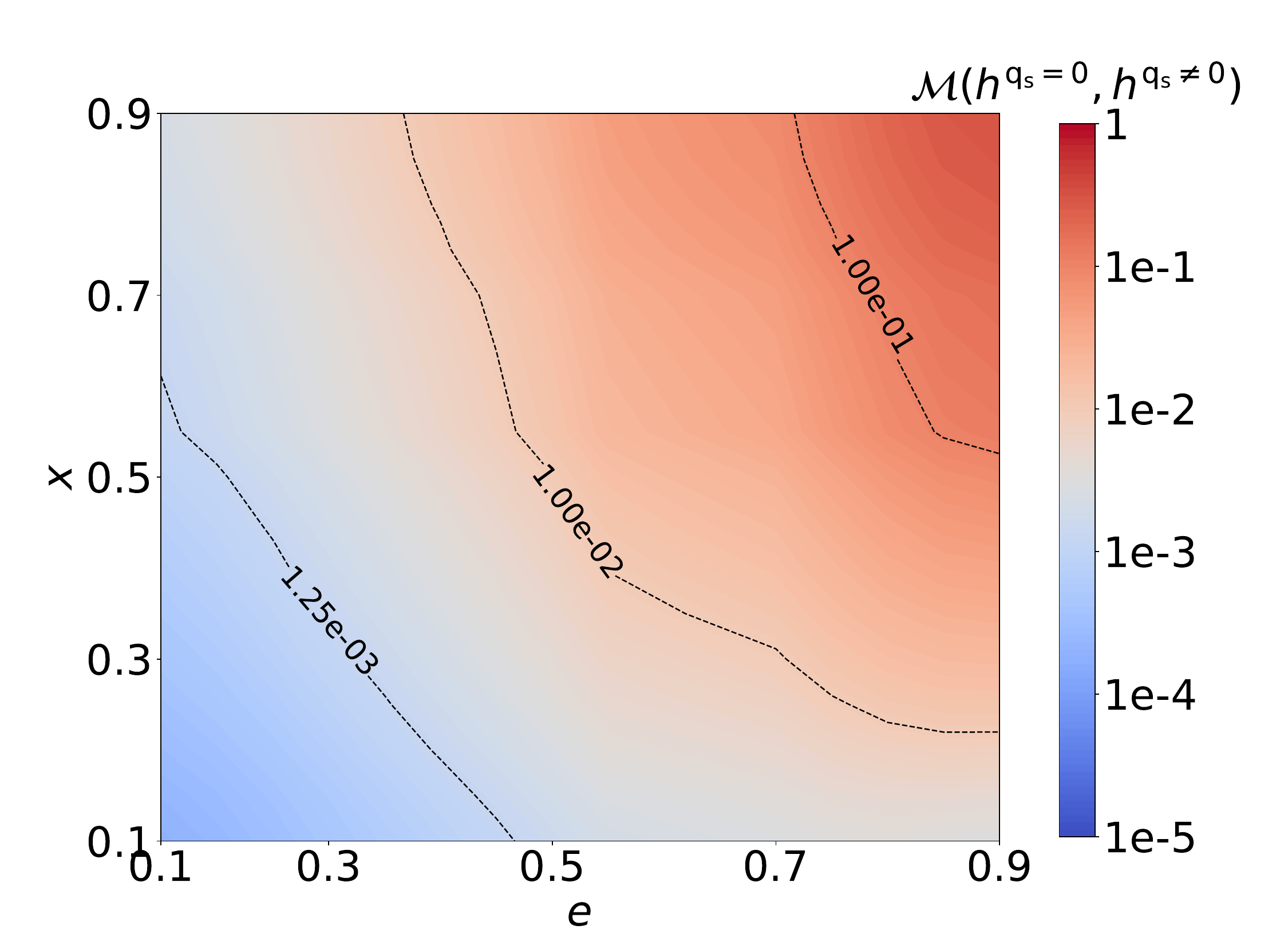}
\includegraphics[width=3.0in, height=2.0in]{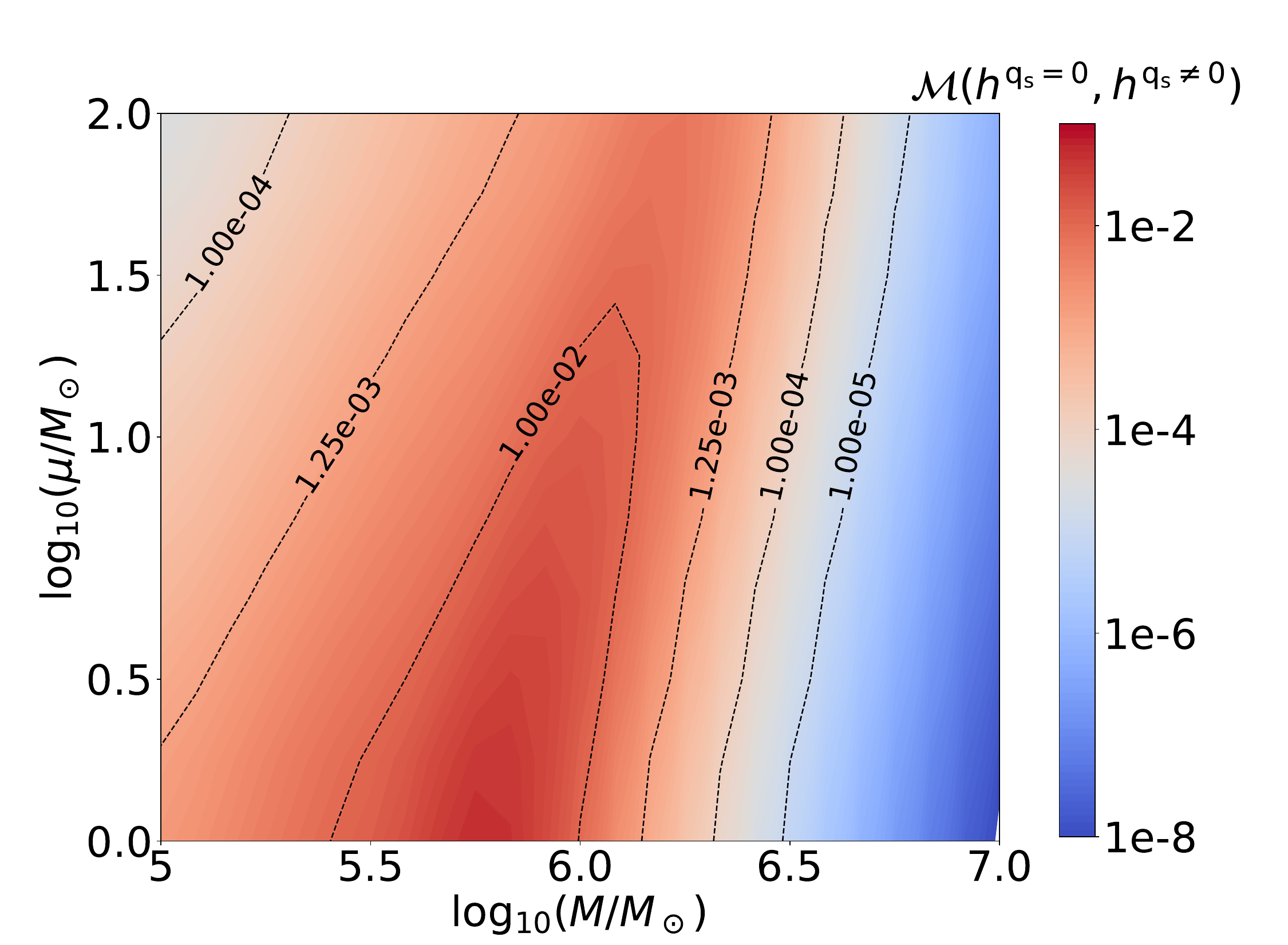}
\includegraphics[width=3.0in, height=2.0in]{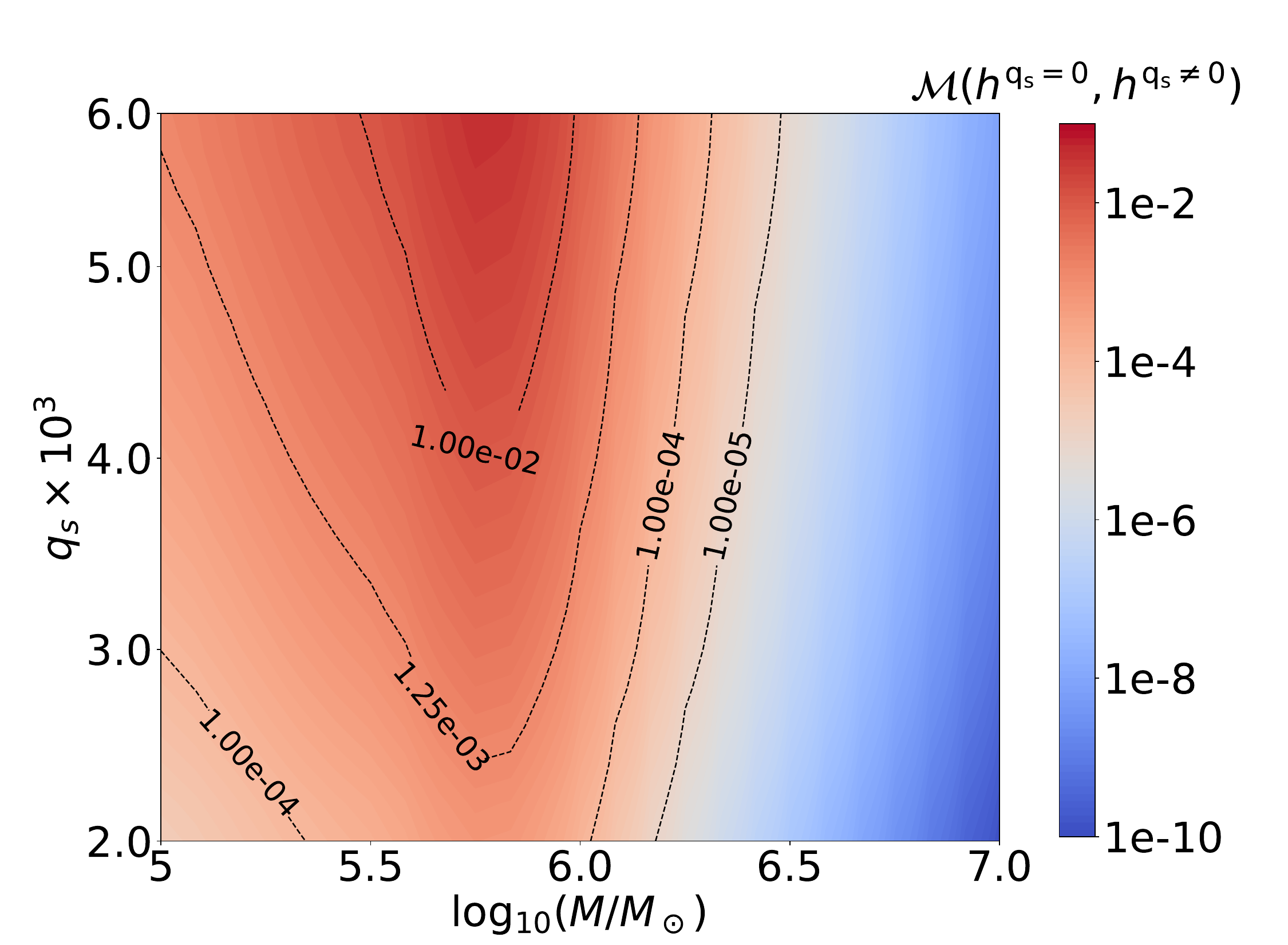}
\caption{Mismatch of EMRI waveforms $(h^{q_s=0}, h^{q_s\neq0})$ with and without the correction of scalar charge as a contour of different parameters setting is plotted, incorporating the dependent relations of $(q_s, e)$ in the top-left panel, $(x, e)$ in the top-right panel, $(\log_{10}(\mu/M_\odot), e)$ in the bottom-left panel and $(q_s, \log_{10}(\mu/M_\odot))$ in the bottom-right panel. The other parameters are set as follows:
the initial semi-latus rectum $p=10$, MBH spin $a=0.3$, the initial orbital inclination $x=0.3$ (top-left panel), scalar charge $q_s=0.004$ (top-right panel), initial parameters $e=0.3, x=0.3$ and scalar charge $q_s=0.005$ (bottom-left panel), initial eccentricity $e=0.3$ and inclination $x=0.3$ (bottom-right panel). 
The mass-ratio in three panels is fixed as $q=10^{-5}$, except the bottom-left panel.} \label{fig:mismatch:a03}
\end{figure*}

In our analysis, the interpolated gravitational and scalar fluxes depend not only on the discrete flux data computed on the sampling grid but also on the accuracy of the interpolation scheme itself. We therefore perform a two-step assessment of interpolation errors. First, we evaluate the local interpolation accuracy by comparing the spline-interpolated fluxes at off-grid points with the corresponding perturbative fluxes. As shown in Fig.~\ref{fig:flux:error}, the interpolation errors for both the scalar energy flux and the scalar Carter flux lie in the range $10^{-6}$--$10^{-4}$, with the largest deviations appearing in the strong-field region.

Second, we examine the cumulative impact of these small local interpolation errors on long-duration inspirals. Following the methodology of Ref.~\cite{Khalvati:2025znb}, we compute the accumulated phase error
\begin{equation}
\Delta \Phi = 
\int_{p_s}^{p_0} 
\omega(a,p,e,x)\,
\frac{E'(a,p,e,x)}
     {\dot{E}(a,p,e,x)}
\,\epsilon_S(a,p,e,x)\, dp,
\end{equation}
where the interpolation deviation is defined as 
$\epsilon_S(a,p,e,x)=\dot{E}_S^{\rm int}-\dot{E}_S^{\rm per}$, 
with $\dot{E}_S^{\rm per}$ denoting the perturbative scalar flux and 
$\dot{E}_S^{\rm int}$ the spline-interpolated value.  
Here $\dot{E}(a,p,e,x)$ is the total (scalar plus gravitational) energy flux, and 
$E'(a,p,e,x)$ is the derivative of the orbital energy with respect to $p$.  
Figure~\ref{fig:local:flux:acc:error} displays the resulting maximum accumulated azimuthal phase error 
$\delta\Psi_\phi^{\rm error,acc}$ for three representative initial eccentricities 
$e_0\in\{0.1,0.3,0.5\}$ as a function of the orbital inclination $x$. 
We find that the accumulated phase error increases with eccentricity, which is likely due to the fact that highly eccentric and inclined EMRI orbits receive significant contributions from a broader set of harmonic modes, as indicated by Eq.~\eqref{fluxes:scal:eqs}, whereas our interpolation framework currently includes up to the dominant-mode contribution. Systematically incorporating higher-mode flux components into the interpolation scheme is therefore a natural direction for future work.

Fig.~\ref{fig:dephasings} shows the azimuthal, radial and polar dephasings as a function of observation time with two orbital eccentricities $e=(0.3,0.5)$, indicating the comparison between the GR and distinct vales of scalar charge $q_s=(10^{-7},10^{-6},10^{-4},10^{-3},10^{-2},10^{-1})$. According to six panels, the dephasing is varying from a tiny fraction to about hundreds of radians. Among those, the azimuthal dephasing is mainly larger than other two other dephasings, which also supports the point in Refs.~\cite{Barsanti:2022ana,Zi:2023geb}.
Comparing two azimuthal dephasings (top-left and middle-right panels), the dephasing is larger for the eccentricity $e=0.5$, for the case of $e=0.3$, the similar conclusions can be seen for the radial and polar dephasings.
If only considering the azimuthal dephasing for orbital eccentricity $e=0.5$, the effect of scalar charge $q_s^{(\min)}\sim10^{-4}$ can be recognized by LISA for the initial orbital inclination $x=0.5$.
This indicates that next-generation detectors can potentially distinguish the effect of scalar charge through EMRI signals. We further perform the mismatch analysis, providing a rigorous computation of examining scalar charge effect and its potentially detectability.

\begin{figure*}[htb!]
\centering
\includegraphics[width=3.2in, height=2.2in]{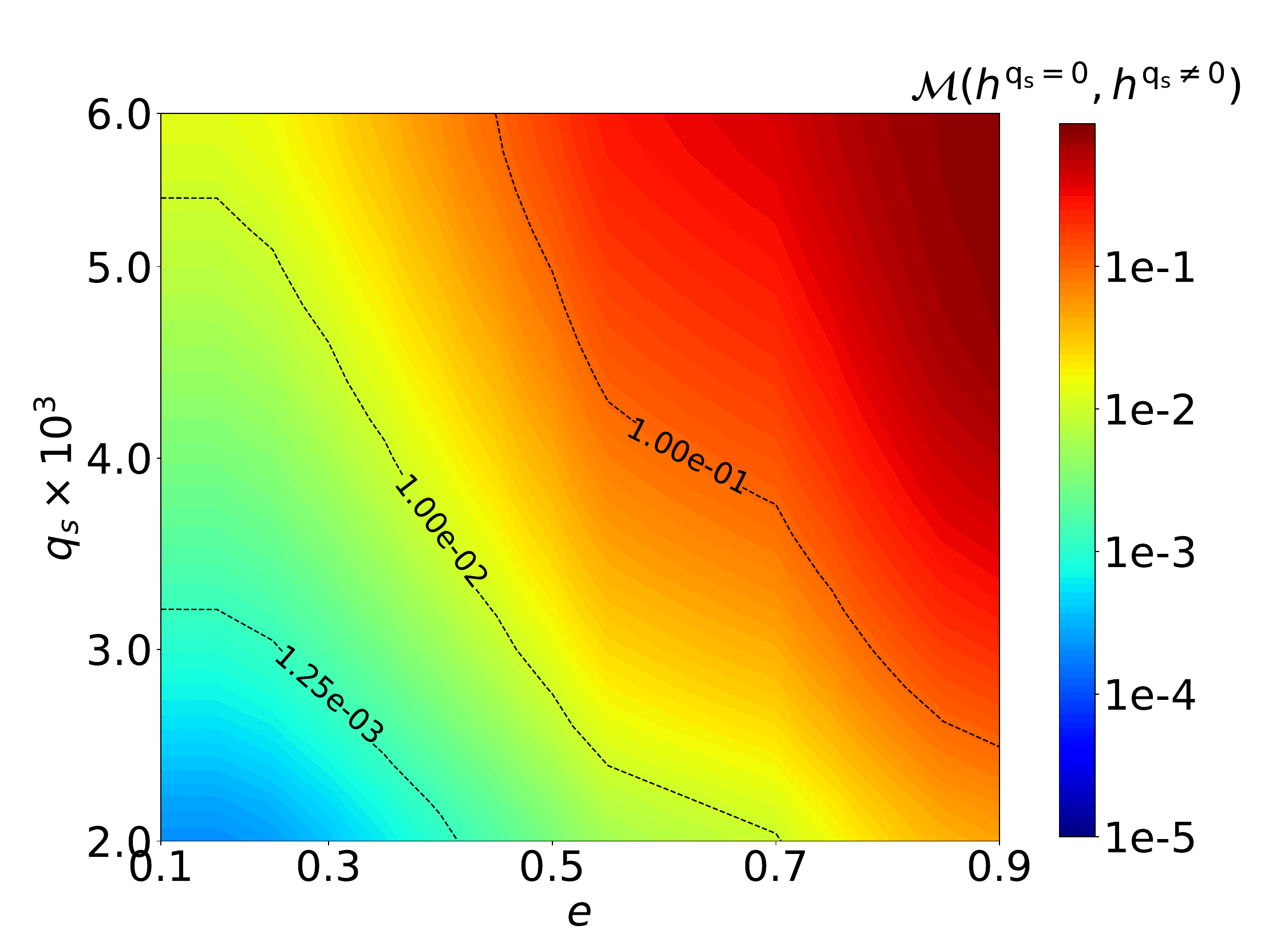}
\includegraphics[width=3.2in, height=2.2in]{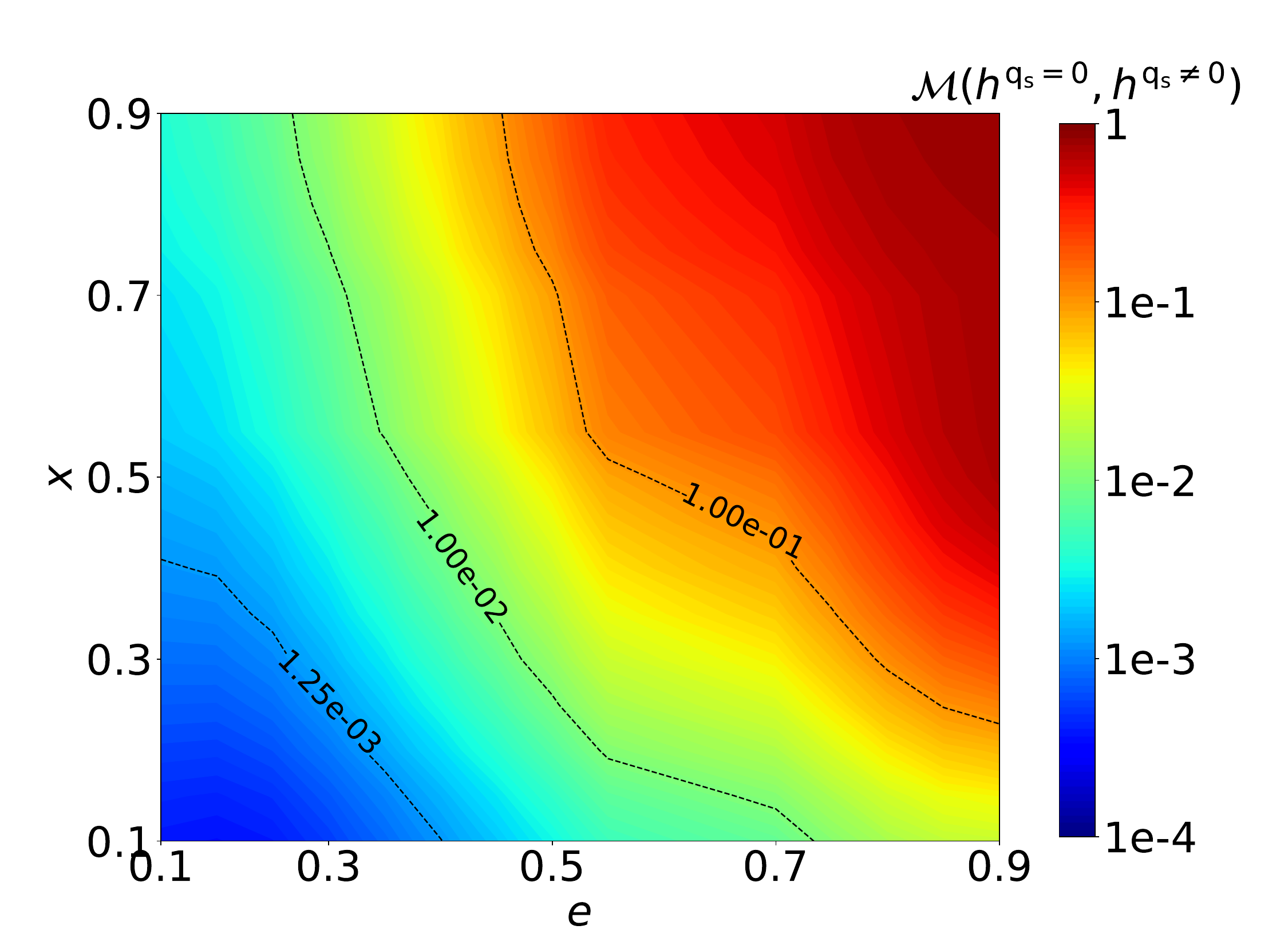}
\includegraphics[width=3.2in, height=2.2in]{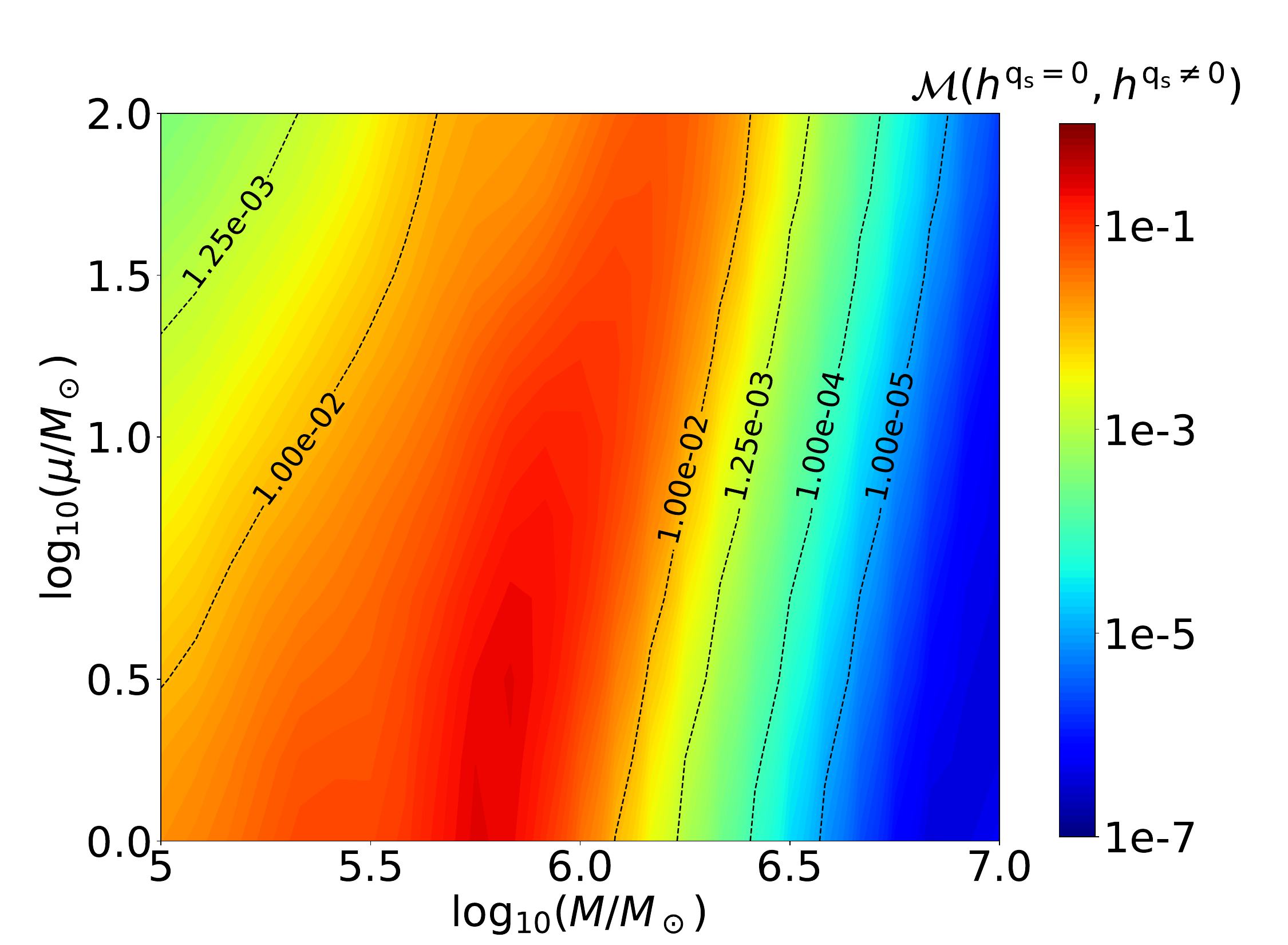}
\includegraphics[width=3.2in, height=2.2in]{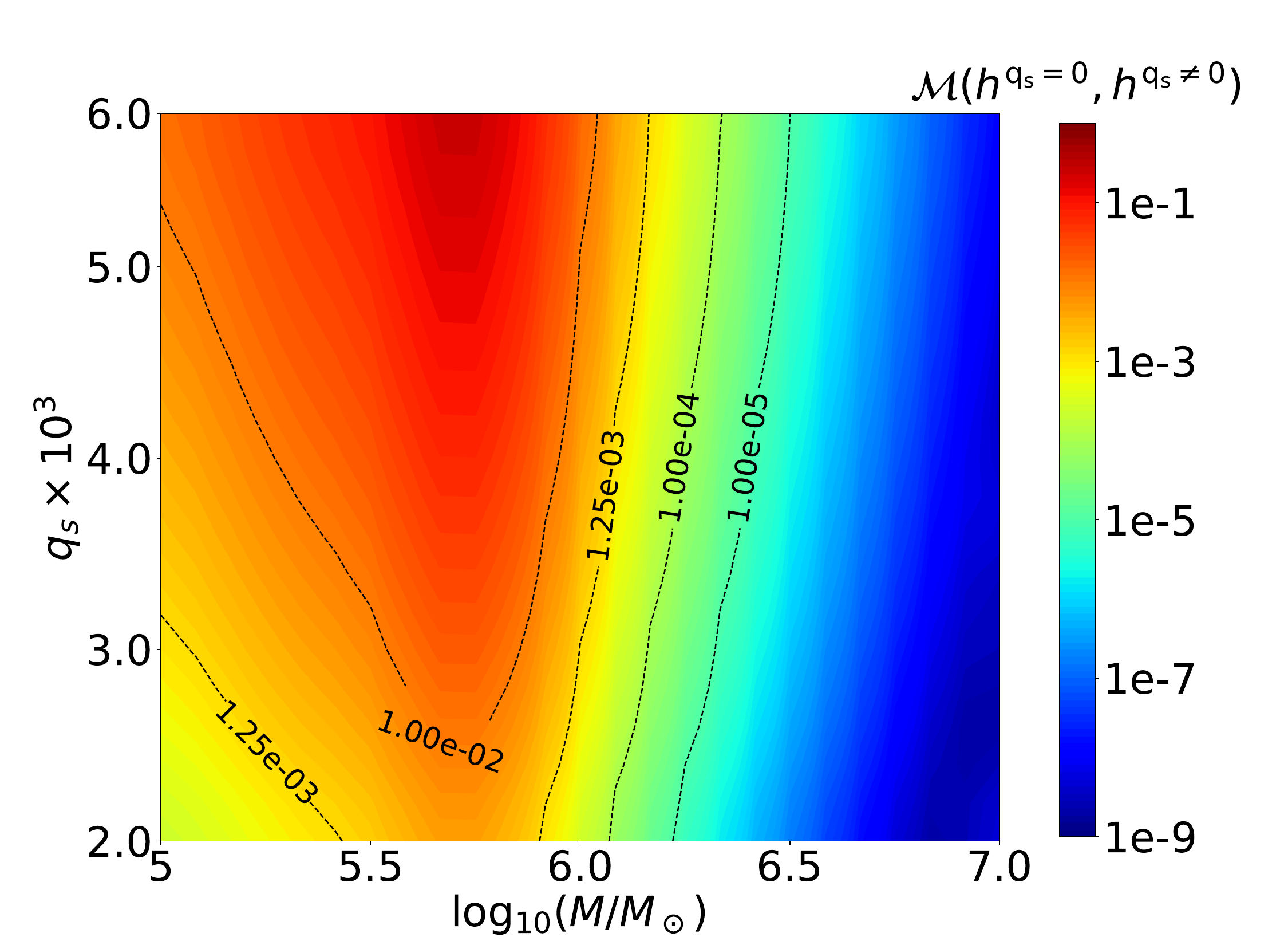}
\caption{Mismatch of EMRI waveforms $(h^{q_s=0}, h^{q_s\neq0})$ with and without the correction of scalar charge as a contour of different parameter settings is plotted for the MBH spin $a=0.9$, incorporating the dependent relations of $(q_s, e)$ in the top-left panel, $(x, e)$ in the top-right panel, $(\log_{10}(\mu/M_\odot), e)$ in the bottom-left panel and $(q_s, \log_{10}(\mu/M_\odot))$ in the bottom-right panel. The values of other parameters are same as in Fig.~\ref{fig:mismatch:a03}.} \label{fig:mismatch:a09}
\end{figure*}
\subsection{Mismatch and constraint on scalar charge}

In Fig.~\ref{fig:mismatch:a03}, we present mismatches $\mathcal{M}(h^{q_s=0}, h^{q_s\neq0})$ as functions of four intrinsic parameters MBH spin $a=0.3$. The black dashed curves denote the contour lines for several mismatch values.
In the top-left panel of Fig.~\ref{fig:mismatch:a03}, we consider the mismatch depending on the scalar charge and eccentricity $(q_s,e)$ for a fixed orbital inclination $x=0.3$. Under these settings of parameters, we find that more eccentric orbits can help us to distinguish the imprints of scalar charge with more pronounced results. The top-right panel shows the mismatch as a function of the orbital inclination and eccentricity $(x,e)$ for a fixed scalar charge $q_s=0.004$. From the top two panels, the increasing orbital inclination is useful to probe the effect of scalar charge using EMRI signals. Further, the distinguishable parameter region in the top-right panel is expanded to compare the case of the top-left panel, when the orbits become more inclined.
Given that scalar charge carried by secondary in the EMRI dynamics, the masses of secondaries may have a significant influence on GW signal. Therefore, the bottom-left panel shows the mismatch as a function of binaries' masses $(\log_{10}(\mu/M_\odot), \log_{10}(M/M_\odot))$ with the initial eccentricity, inclination $(e=0.3,x=0.3)$ and scalar charge $q_s=0.005$. We notice that the signal from EMRIs with the primary's mass $M\gsim10^{6.2}M_\odot$ can not be detected regardless of the masses of secondaries. Importantly, there is a part of the mass-ratio areas (MBH with a mass $M\lsim10^{6.2}M_\odot$ and mass-ratio $q\in[10^{-6},10^{-4}]$) where the effect of the scalar charge can be discerned by LISA, so we should be cautious of masses of binary systems as they can influence the relative magnitude of scalar-charge effects when comparing beyond-GR EMRI waveforms with their vacuum Kerr counterparts.
In the bottom-right panel, we further investigate how the scalar charge and primary's mass $(q_s,\log_{10}(M/M_\odot))$ impacts the mismatch of two GW signals for initial eccentricity, inclination $(e=0.3,x=0.3)$ and a fixed smaller object's mass $\mu=10~M_\odot$.
We notice that LISA can distinguish the effects of the scalar charge as small as a fraction of $q_s\sim2.5\times10^{-3}$.

In Fig.~\ref{fig:mismatch:a09}, we also consider four cases of mismatch computation for the higher spins of MBH $a=0.9$, keeping the other parameters the same as mentioned in Fig.~\ref{fig:mismatch:a03}. It is worth emphasizing that scalar charge-corrected EMRI signals around rapidly spinning MBHs are more significantly distinguishable in the two-dimensional parameter space when other intrinsic parameters are held fixed, as shown in Fig.~\ref{fig:mismatch:a03}. For example, in the top-right panels of Fig.~\ref{fig:mismatch:a03} and Fig.~\ref{fig:mismatch:a09}, for a scalar charge of $q_s = 0.004$, the minimum values of the orbital parameters discernible by LISA are approximately $(x,e)_{\min} \sim (0.6, 0.48)$ for a spin of $a = 0.3$, which decrease to $(x,e)_{\min} \sim (0.4, 0.4)$ in the case of a higher spin, $a = 0.9$. This implies that the highly spinning BHs indeed magnify/enhance the effect of scalar charge; it can possibly be due to the fact that secondaries moving around the MBH can accumulate more orbital cycles and consequently the EMRI signal from such a highly spinning MBH gives rise to more rich information of scalar emission. Compared to the top-right panels in Fig.~\ref{fig:mismatch:a03} and Fig.~\ref{fig:mismatch:a09}, the remaining three panels more clearly illustrate the expansion of the parameter space that LISA can distinguish.

\begin{table*}[htbp!]
\centering
\begin{tabular}{cc|ccccccccccccc}
\hline
\hline
$q_s$ & $x$ & $\sigma_M/M$ & $\sigma_{\mu}/\mu$ &$\sigma_{a}$   &$\sigma_{p}$   & $\sigma_{e}$   &$x$   & $\sigma_{q_s}/q_s$
& $\sigma_{\Phi{_{\phi,0}}}/\Phi_{\phi,0}$    & $\sigma_{\Phi{_{r,0}}}/\Phi_{r,0}$ 
& $\sigma_{\Phi{_{\theta,0}}}/\Phi_{\theta,0}$  & $\sigma_{d_L}$
\\
\hline
0.01  & 0.1
&$1.24\text{e-5}$    &$3.99\text{e-4}$    &$5.29\text{e-4}$
&$2.33\text{e-4}$  & $4.33\text{e-4}$    & $3.59\text{e-5}$
& $7.98\text{e-2}$   & $6.36\text{e-1}$
&$7.57\text{e-1}$ &$4.57\text{-1}$      &$8.24$
\\
&0.3 
&$4.53\text{e-5}$  &$5.67\text{e-4}$ &$5.45\text{e-4}$   &$3.47\text{e-4}$ 
&$6.26\text{e-4}$  &$1.43\text{e-4}$  &$8.16\text{-2}$  &$6.76\text{e-1}$
&$5.37\text{e-1}$ 
&$7.48\text{e-1}$    &$5.25$
\\
&$0.5$   
&$5.67\text{e-5}$  &$6.75\text{e-4}$  &$6.05\text{e-4}$   &$6.78\text{e-4}$ 
&$5.62\text{e-4}$   &$7.89\text{e-4}$  &$7.38\text{e-2}$ &$3.56\text{e-1}$  &$7.12\text{e-1}$  &$1.48\text{e-1}$    &$6.42$
\\
\hline
\hline
0.03  & 0.1  
&$2.12\text{e-5}$  &$4.16\text{e-4}$  &$4.36\text{e-4}$ &$3.57\text{e-4}$  & $4.24\text{e-4}$    & $2.14\text{e-5}$
& $6.43\text{e-2}$   & $5.24\text{e-1}$
&$7.24\text{e-1}$ &$4.12\text{-1}$      &$7.35$
\\
&0.3   &$3.27\text{e-5}$  &$5.24\text{e-4}$ &$5.57\text{e-4}$   &$4.27\text{e-4}$   & $7.14\text{e-4}$ 
&$2.34\text{e-4}$  &$7.37\text{-2}$  &$8.15\text{e-1}$
&$6.45\text{e-1}$ 
&$5.76\text{e-1}$    &$6.15$
\\
&0.5 
&$4.26\text{e-5}$  &$5.42\text{e-4}$  &$5.45\text{e-4}$   &$4.57\text{e-4}$
&$5.62\text{e-4}$   &$3.46\text{e-4}$  &$7.43\text{e-2}$
&$4.21\text{e-1}$  &$6.14\text{e-1}$ 
 &$5.45\text{e-1}$    &$5.67$
 \\
\hline
\hline
\end{tabular}
\caption{ Measurement errors for EMRIs intrinsic parameters, masses $(M=10^{6}M_\odot, \mu = 10M_\odot)$,
MBH's spin $a=0.9$, orbital semi-latus rectum and eccentricity $(p=10,e=0.3)$,  initial orbital phases $(\Phi_{\phi,0}=1.0)$ and
the directional angles related to source $(\theta_{S,K}=\phi_{S,K}=1.0)$ are listed, where the secondary object lasts the spiraling of two years and the luminosity distance $(d_L)$ is adjusted to set SNR of EMRIs signal as $50$.
}\label{tab:fim:error}
\end{table*}

\begin{figure*}[htb!]
\centering
\includegraphics[width=0.88\paperwidth]{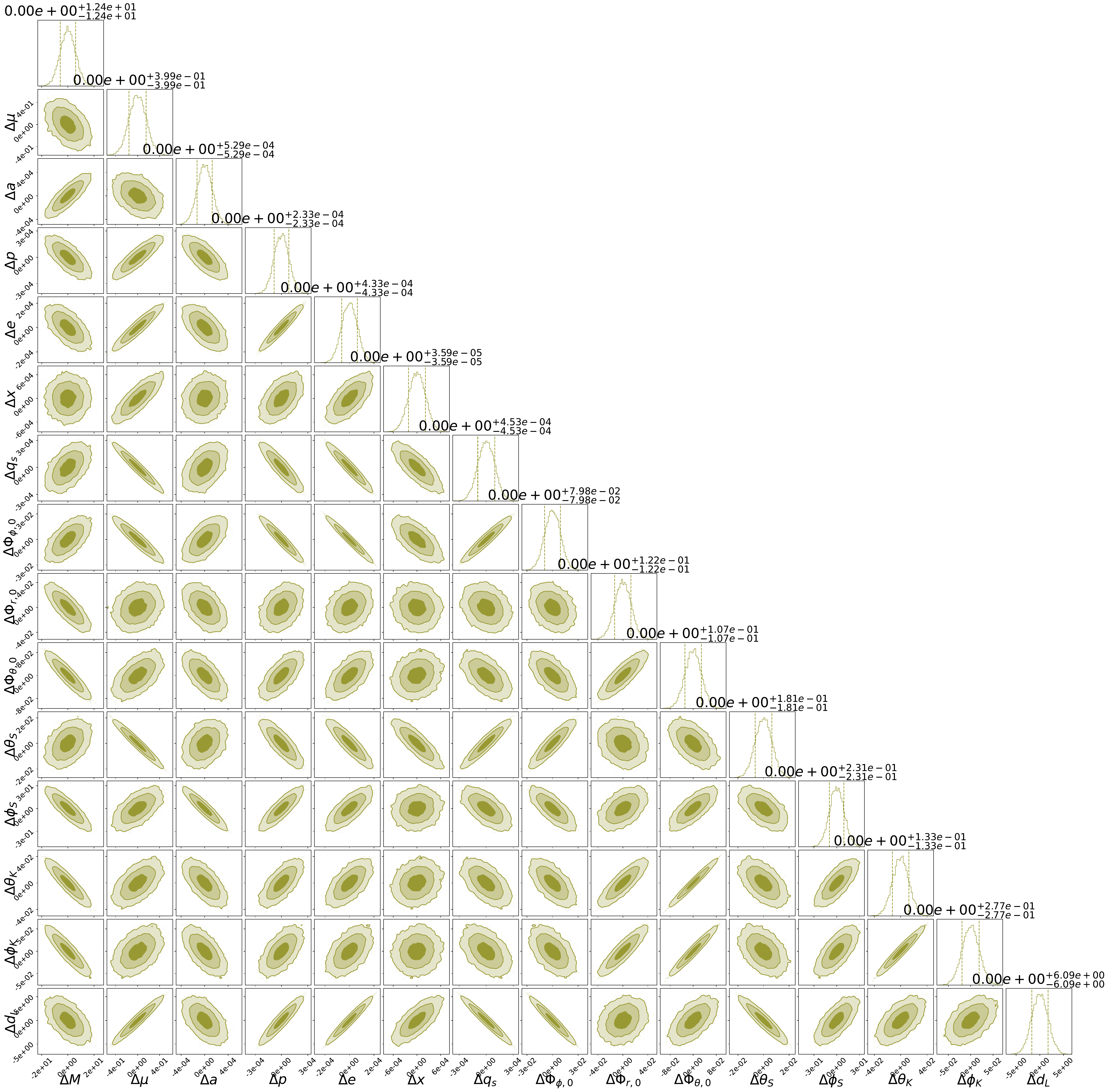}
\caption{Corner plot of posterior probability densities for EMRIs embedded in a Hernquist-type dark matter environment. We assume $(M=10^{6}M_\odot,\; \mu=10M_\odot,\;a=0.9, p=10, e=0.3, x=0.1, \; \Phi_{\phi,0}=\Phi_{r,0}=\Phi_{\theta,0}=1.0,\; q_s=0.01\; $. 
All extrinsic parameters are fixed as in \autoref{tab:fim:error}. 
Posteriors are inferred from a two-year LISA observation. Vertical dashed lines denote the $1\sigma$ credible intervals; shaded contours show the $68\%$, $95\%$, and $99\%$ credible regions.} \label{fig:cornerplot:Hern}
\end{figure*}

\begin{figure*}[htb!]
\centering
\includegraphics[width=4.85in, height=4.15in]{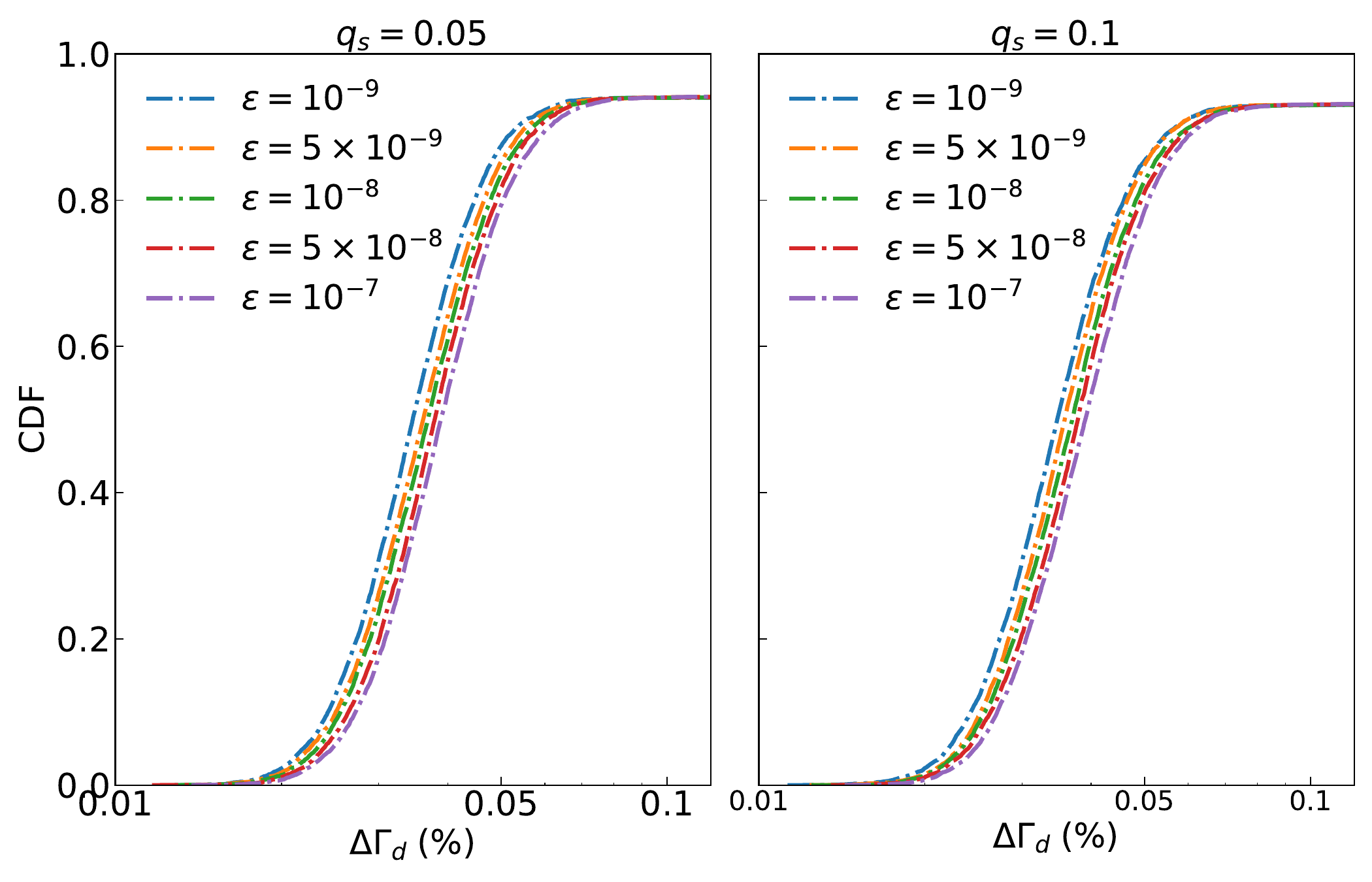}
\caption{Cumulative distribution function (CDF) for the maximum relative error between the perturbed covariance matrices is plotted for two scalar charges 
$q_s\in\{0.01,0.1\}$ and other initial orbital parameters $(p_0=10,e_0=0.3, x=0.3)$. Different colored curves denote to the FIM computed with a different numerical derivative spacing for scalar charge, the elements of perturbed  matrix are defined in the range of $[-10^{-3},10^{-3}]$. } \label{fig:CDF:qs}
\end{figure*}

To present a preliminary constraint on scalar charge with a generic EMRI signal in LISA bank, we compute the parameter estimation for all source's parameters using FIM. Because the evolution of EMRI waveform phase is mainly governed by the intrinsic parameters, we only list the measurement error of these intrinsic parameters in Table~\ref{tab:fim:error}, outlining the measurement error of intrinsic parameters, including two cases of scalar charge $q_s=0.01$ and $q_s=0.03$ and three case of orbital inclination $x\in\{0.1,0.3,0.5\}$. One can find that there is slightly improvement of measurement error of scalar charge when the charge $q_s$ is bigger, the orbital inclination do not influence the relative measurement error $\sigma_{q_s}/q_s$ significantly. The orbital configuration with $x = 0.1$ ($\sim 0.47\pi$) imposes more stringent constraints on these parameters, it may be because the orbits on the near equatorial plane can carry more information on the strong-filed region of MBH.
In numerical analysis of FIM, the diag-element of the covariance matrix can approximately give the measurement error for each parameter, the off-diag elements can determine the correlation of source parameters. So we here perform a correlation analysis for all source parameters in Fig.~\ref{fig:cornerplot:Hern}, considering the case of scalar charge $q_s=0.01$. We find that there is a negative correlation between the scalar charge $q_s$ and other parameters $(\mu, p,e,x)$; however the correlation for parameters $q_s$ and primary object's mass $M$ is positive, and the correlations between the $q_s$ and extrinsic parameters are also positive. These correlations between scalar charge and other parameters suggest the possibility of constraining such effects in future with high SNR EMRI observations.

Finally, to assess the robustness of the FIM and its inverse, we adopt the procedure outlined in Refs.~\cite{Vallisneri:2007ev, Zi:2022hcc, Speri:2021psr, Zi:2025onl}. In particular, we examine the numerical stability of the inversion process for a FIM of dimension $15 \times 15$. We begin by constructing a perturbation matrix $\mathbf{R}$ of the same size as the FIM $(\mathbf{\Gamma})$, where each element of $\mathbf{R}$ is drawn from a uniform distribution $u \in [-10^{-3}, 10^{-3}]$. The perturbed inverse matrix $(\mathbf{R} + \mathbf{\Gamma})^{-1}$ is then computed, and we quantify the numerical stability by evaluating the maximum relative deviation between the perturbed and unperturbed covariance matrices. Specifically, we define
\begin{equation}\label{eq:relative:dev}
\delta \mathbf{\Gamma}_d \equiv \max \left( 
\frac{(\mathbf{R}+\mathbf{\Gamma})^{-1}-\mathbf{\Gamma}^{-1}}{\mathbf{\Gamma}^{-1}} 
\right).
\end{equation}
The distribution of $\delta \mathbf{\Gamma}_d$ provides a direct measure of the sensitivity of the inversion to small perturbations in the FIM. In Fig.~\ref{fig:CDF:qs}, we present the cumulative distribution of $\delta \mathbf{\Gamma}_d$ for different values of the scalar charge $q_s$ and setting other initial orbital parameters $(p_0=10,e_0=0.3, x=0.3)$, computed using five derivative spacings $\epsilon \in \{10^{-9},\, 5\times10^{-9},\, 10^{-8},\, 5\times10^{-8},\, 10^{-7}\}$. The results show that more than $90\%$ of the population yield a relative deviation $\Delta \Gamma_d \lesssim 0.05\%$, indicating that the FIM inversion is numerically stable with respect to variations in $q_s$. Similar levels of stability are observed for the remaining parameters, confirming the overall reliability of the FIM computation.

\section{Discussion}\label{dscn}
EMRIs are one of the most promising target sources for next-generation space-based detectors such as LISA, offering an unparalleled opportunity to test GR in the strong-field regime. These sources serve also as probes of beyond-GR effects due to their extreme sensitivity to small perturbations through inspiral dynamics, reflected in observables, including those arising from additional radiation channels such as scalar emission, even when the background spacetime remains effectively Kerr.
Accurate modeling of scalar field effects in EMRIs is therefore crucial for robust parameter estimation, enabling precise extraction of source properties and the potential identification of additional radiation channels beyond vacuum GR dynamics, i.e., scalar-induced effects in our context.

Recently, computing EMRIs beyond GR has made significant progress in presenting the constraint on scalar charge in the framework of perturbation theory, where these corrections are reflected in the fluxes emitted from the circular, inclined and eccentric insprialling orbits~\cite{Maselli:2020zgv,Maselli:2021men,Barsanti:2022ana,Barsanti:2022vvl,Burke:2023lno,DellaRocca:2024pnm,Speri:2024qak}. In this paper, we first compute the modifications of the scalar charge on generic EMRI fluxes, the gravitational and scalar fluxes can be computed over a three-dimensional grid $(e,p-p_{\rm sp},x)$ to generate fluxes efficiently using the interpolation method, enabling us to evolve the orbital parameters adiabatically.
By evolving the inspirals up to the merger for cases with and without a scalar charge, the impact of the scalar charge $q_s$ on EMRI waveforms can be evaluated through the computation of dephasing and mismatch.

Based on the comparison of energy and Carter constant fluxes from gravitational and scalar emissions in Fig.~\ref{fig:flux:ratio}, the contribution of scalar flux to EMRI evolution diminishes as the orbital inclination $x$ increases, for a fixed initial eccentricity. Moreover, the ratio of the scalar and gravitational fluxes is non-monotonically varying with larger eccentricity and for a fixed orbital inclination $x$. To verify the feasibility of interpolation fluxes over the sampling grids,
we plot the different values of fluxes from the interpolation method off the sampling grids in Fig.~\ref{fig:flux:error}, where the errors are mainly in the range of $[10^{-6},10^{-4}]$. 
Figure~\ref{fig:local:flux:acc:error} illustrates the accumulated phase error induced by small local interpolation inaccuracies over long-duration inspirals. Our analysis of the interpolation scheme suggests that the fluxes, consequently in dephasing with distinct mass-ratios, obtained with the current approach may introduce a non-trivial and potentially non-negligible error at higher eccentricities as presented in Table~\ref{tab:interpolation:mis}. Nevertheless, we expect that the method remains sufficient for a qualitative investigation of scalar charge effects on EMRI dynamics and can still be used to approximately model the adiabatic evolution of generic geodesic orbits.
In Fig.~\ref{fig:dephasings}, we further compute the dephasings of inspiraling orbits using the fluxes with and without the correction of scalar charge for the eccentricities $e=(0.3,0.5)$, setting the spin of MBH $a=(0.3,0.5)$. We find that the azimuthal dephasing is larger than the other dephasings for the radial and polar directions. LISA can discern the effect of the scalar charge as small as $10^{-3}$ for a fixed initial eccentricity and inclination parameter $(e=0.3,x=0.5)$. Consequently, such an effect with smaller values of the scalar charge, which becomes significant, can potentially be detected for eccentric orbits.
Finally, we assess the mismatches influenced by different parameters for EMRI waveforms within the standard GR and the modified gravity in Fig.~\ref{fig:mismatch:a03}.
Our mismatch analysis indicates the following points: the effect of scalar charge becomes increasingly distinguishable by LISA in EMRIs with higher orbital eccentricities; the threshold scalar charge detectable by LISA decreases as the orbital inclination parameter $x$ increases; the scalar charge carried by EMRIs can be detected when the mass of primary satisfies $M\lsim 10^{6.2}M_\odot$ and mass-ratio $q\in[10^{-6},10^{-4}]$.
When the primary object is a rapidly rotating ($a=0.9$) MBH, as shown in Fig.~\ref{fig:mismatch:a09}, the distinguishable parameter region for detecting the effects of the scalar charge expands significantly compared to the lower spin case ($a=0.3$) shown in Fig.~\ref{fig:mismatch:a03}.
Finally, we present constraints on the scalar charge obtained from EMRI signals with generic inspiral orbits using the FIM approach. Our analysis indicates that the relative measurement uncertainty of the scalar charge can be constrained to a fractional level of $\sigma_{q_s}/q_s \sim 10^{-2}$ with LISA observations, as shown in Table~\ref{tab:fim:error}. We find that the precision of EMRI parameter estimation improves as the orbital inclination $x$ decreases. Furthermore, by examining the off-diagonal elements of the covariance matrix, we assess the correlations among all source parameters. The scalar charge $q_s$ exhibits negative correlations with the secondary mass $\mu$, semi-latus rectum $p$, eccentricity $e$, and inclination $x$, while it shows positive correlations with the primary mass $M$ and with the extrinsic parameters. By introducing a perturbation matrix to the FIM, we confirm that the resulting covariance matrix remains stable for scalar charge.

In our analysis, dephasing, mismatch and FIM provide only a preliminary estimate of the scalar charge's impact on EMRI waveforms beyond GR. However, these measures do not capture the correlations among different parameters within the full source parameter space. It is essential to implement a fully Bayesian analysis based on Monte Carlo Markov Chain simulations with EMRI Waveforms with the correction of scalar charge~\cite{Katz:2021yft,Speri:2024qak,Khalvati:2024tzz,Chapman-Bird:2025xtd}.
Post-adiabatic corrections to scalar and gravitational fluxes are important for improving the accuracy of EMRI waveforms \cite{Spiers:2023cva}. Further, the effect of a spinning secondary on scalar emission has also gained interest as a potential source of new observable signatures \cite{Burke:2023lno}. We aim to report some of these studies in future communications.

\section*{Acknowledgements} 
T.Z. is funded by the National Natural Science Foundation of China with Grant No. 12347140 and No. 12405059. The research of S.K. is funded by the National Post-Doctoral Fellowship (N-PDF: PDF/2023/000369) from the ANRF (formerly SERB), Department of Science and Technology (DST), Government of India. S.K. also extends sincere thanks to the organizers of Capra 28, and to Leor Barack and Adam Pound for facilitating the post-Capra visit during the progress of this work.
\appendix 
\section{Source term for gravitational perturbation}\label{app:source:grav}
The source term in the Teukolsky perturbation equation in the background of Kerr BH has been derived in Refs.~\cite{Teukolsky:1972my,Teukolsky:1973ha,Hughes:1999bq,Drasco:2005kz}
\begin{equation}\label{source_term}
\mathcal {T} _ {\ell m \omega} =  4 \int dt d\theta\sin\theta d\phi \frac{\left(B' _ 2 + {B' _ 2}^*\right)}{\bar{\rho}\rho^5} S_{\ell m\omega} e^{-  i (m\phi+ \omega t)}
\end{equation}
where,
\begin{align}
B' _ 2 =& - \frac{1}{2} \rho^8\bar {\rho}\mathcal {L} _{-1} 
\bigg[\frac{1}{\rho^4}\mathcal{L}_0\bigg[\frac{T_{nn}}{\rho^2\bar{\rho}} \bigg]\bigg]
 \\ &-\frac{1}{2\sqrt{2}}\Delta^2 \rho^8\bar{\rho}\mathcal{L}_ {-1}\bigg[\frac{\bar{\rho}^2}{\rho^4} 
J_+\bigg[\frac{T_{\overline{m}n}}{ \hat{\Delta} \rho^2\bar{\rho}^2} \bigg]\bigg] \ , \\
 {B' _ 2}^*=& - \frac {1} {4}\Delta^2 \rho^8\bar{\rho} J_+\bigg[\frac{1}{\rho^4}J_+ 
\bigg[\frac{\bar{\rho}}{\rho^2}T_{\overline{m}\overline{m}}\bigg] \bigg]
\\ & - \frac{1}{2\sqrt {2}}\Delta^2 \rho^8\bar{\rho} J_+ \bigg[\frac{\bar{\rho}^2}{\Delta \rho^4}\mathcal {L}_ 
{-1}\bigg[\frac{ T_ {\overline{m}n}}{\rho^2\bar {\rho}^2}\bigg] \bigg] \ ,
\end{align}
where we remind that $\Delta = r^{2}-2Mr+a^{2}$, $K=(r^{2}+a^2)\omega-ma$, $\rho = 1/(r-i a\cos\theta) = -\bar{\rho}$ and operators.
\begin{equation}
\begin{aligned} 
J_+ =& \frac{\partial}{\partial r}+\frac{i K}{\Delta} \hspace{0.1cm} ; 
\hspace{0.1cm}\mathcal{L} _s = \frac{\partial}{\partial\theta}+\frac{m}{\sin \theta}+s \cot\theta - a\cot\theta\; \\ 
\mathcal{L} _s^\dagger =& \frac{\partial}{\partial\theta}-\frac{m}{\sin \theta} + s 
\cot\theta - a\cot\theta \;. 
\end{aligned}
\end{equation}
Quantities ($T_{nn}, T_{\bar{m}n}, T_{\bar{m}\bar{m}}, T_{\bar{m}n}$) represent the projections of the stress-energy tensor onto the Newman-Penrose (NP) tetrad basis. Following \cite{Sasaki:2003xr, PhysRevD.102.024041, Zi:2024dpi}, we can write down the stress-energy tensor of the point particle as
\begin{align}\label{source1}
T^{\mu\nu} = \frac{m_p(dt/d\tau)^{-1}}{\Sigma \sin\theta}\frac{dz^{\mu}}{d\tau}\frac{dz^{\nu}}{d\tau}
\delta[r-r(t)]\delta[\theta -\theta(t)]\delta[\phi -\phi(t)],
\end{align}
where $z^{\mu} = (t, r(t), \theta(t), \phi(t))$ is the geodesic trajectory with the proper time $\tau = \tau(t)$. Eq. (\ref{source1}) uses geodesic velocities, which can be replaced using Eq. (\ref{geodesic}). Thus the tetrad components of the stress-energy tensor are given by:
\begin{equation}
\begin{aligned}
T_{nn} =& \frac{m_p C_{nn}}{\sin\theta}\delta[r-r(t)]\delta[\theta-\theta(t)]\delta[\phi-\phi(t)], \\
T_{\bar{m}n} =& \frac{m_p C_{\bar{m}n}}{\sin\theta}\delta[r-r(t)]\delta[\theta-\theta(t)]\delta[\phi-\phi(t)], \\
T_{\bar{m}\bar{m}} =& \frac{m_p C_{\bar{m}\bar{m}}}{\sin\theta}\delta[r-r(t)]\delta[\theta-\theta(t)]\delta[\phi-\phi(t)],
\end{aligned}
\end{equation}
where
\begin{equation}
\begin{aligned}
C_{nn} =& \frac{1}{4\Sigma^{2}}\Big(\frac{dt}{d\lambda}\Big)^{-1} \Big[E(r^{2}+a^2)+\frac{dr}{d\tau} \Big]^{2}, \\
C_{\bar{m}n} =& \frac{\rho}{2\sqrt{2}\Sigma}\Big(\frac{dt}{d\lambda}\Big)^{-1} \Big[E(r^{2}+a^2)+\frac{dr}{d\lambda} \Big]
\\ &\times\Big[i(a E \sin\theta-L_{z}\csc\theta)+\frac{d\theta}{d\lambda}\Big], \\
C_{\bar{m}\bar{m}} =& \frac{\rho^{2}}{2\Sigma}\Big(\frac{dt}{d\lambda}\Big)^{-1}\Big[i(aE \sin\theta-L_{z}\csc\theta)+\Sigma\frac{d\theta}{d\lambda} \Big]^{2}.
\end{aligned}\label{eq:Cnms}
\end{equation}
We note that the Eqs.~\eqref{eq:Cnms} are related with the geodesic equations $dr/d\lambda, ~dt/d\lambda$ and $d\theta/d\lambda$.
The amplitudes $Z^{H,\infty}_{\ell m\omega}$ in Eqs.~\eqref{ampl:grav} can be simplified using the source term and Eq.~\eqref{source_term}
\begin{equation}
Z^{H,\infty}(r,\omega) = \int dt e^{i(\omega t -\phi(t))} I^{H,\infty}_{lm}(t,r,\omega)
\end{equation}
where 
\begin{equation}
\begin{aligned}\label{I:source:grav}  
I^{H,\infty}(r,\omega) &= \Theta[r,r(t)] \Big[ (A_{nn0} + A_{\bar{m}n0}+A_{\bar{m}\bar{m}0}) 
\\  &- (A_{\bar{m}n1}+A_{\bar{m}\bar{m}1})\frac{d}{dr} + A_{\bar{m}\bar{m}2}\frac{d^2}{dr^2}\Big]R^{H,\infty}_{\ell m\omega}
\end{aligned}
\end{equation}
The functions $A_{nna}, A_{\bar{m}nb}, A_{\bar{m}\bar{m}c}$ with $a=0, b\in\{0,1\},c\in\{0,1,2\}$ depend on the quantities $(r, \theta,dr/d\lambda, d\theta/d\lambda)$, their full expressions can be referred to Refs.~\cite{Drasco:2005kz,Sago:2005fn,Isoyama:2018sib}.
$\Theta[r,r(t)]$ is the Heaviside function.
As the radial and polar are periodic functions, which are rewritten as $\theta(w_\theta\equiv\Upsilon_\theta \lambda),  r(w_r\equiv\Upsilon_r \lambda)$, the functions $Z^{H,\infty}_{\ell m\omega}$ can finally be reduced as
\begin{equation}
\begin{aligned}
Z^{H,\infty}_{\ell mkn} = & \frac{1}{2\pi\Gamma}\int_0^{2\pi}dw_\theta \int_0^{2\pi} dw_r e^{ikw_\theta+nw_r} 
I^{H,\infty}_{\ell m}(w_r,w_\theta, r, \omega_{mkn}) ,\\
= &\frac{1}{2\pi\Gamma}\int_0^{2\pi}d\chi \int_0^{2\pi} d\varphi \frac{dw_r}{d\varphi}\frac{dw_\theta}{d\chi}
\\& \times \frac{dt}{d\lambda}
I^{H,\infty}_{\ell m}(w_r,w_\theta, r, \omega_{mkn}) 
e^{i[\omega_{mkn}t(\varphi,\chi)-m\phi(\varphi,\chi)]} 
\;,
\end{aligned}
\end{equation}
where the full expressions of $\frac{dw_\theta}{d\chi}$ and $\frac{dw_r}{d\varphi}$ can be found in Ref.~\cite{Drasco:2005is}.
The computation of coefficients $Z^{H,\infty}_{\ell mkn}$ are integrated over the parameters $(w_r,w_\theta)$, 
in essence,  we can replace the parametrization in Eq.~\eqref{eq:rpsi} and Eq.~\eqref{eq:zchi} to perform the integration over $(\varphi,\chi)$ for the generic orbits. 
The optimization numerical method for computing amplitudes has been argued in Refs.~\cite{Drasco:2005kz,Hughes:2021exa}.


\section{Source term for scalar perturbation}\label{app:source:scal}
Recently, the works have derived the source term and scalar fluxes for inclined circular \cite{DellaRocca:2024pnm}  and eccentric~\cite{Barsanti:2022ana} orbits.
In the section, we briefly introduce the source term and amplitudes for the scalar perturbation.
The source term for a moving point-particle with a scalar charge can be given by
\begin{equation}
\begin{aligned}    
T^s_{\ell mkn}(r,\omega)  =& -4\pi q_s m_p \delta[r-r(t)]  \delta[\phi-\phi(t)] \delta[\theta-\theta(t)]
\\ & \times \left(\frac{dt}{d\lambda}\right)^{-1}S^{s=0}_{\ell kmn}(\theta)\;.
\end{aligned}
\end{equation}
The amplitudes for scalar field in Eq.~\eqref{ampls:scal} begin to written down
\begin{equation}
\begin{aligned}\label{eq:Zs:ampl}
Z^{s,\pm}_{\ell m\omega} = \int_{-\infty}^\infty 
~I^{s,\pm}[r(t),\theta(t)]  e^{i[\omega t -m \phi(t)]} dt 
\end{aligned}
\end{equation}
where 
\begin{equation}
I^{s,\pm}[r(t),\theta(t)] =-\Bigg[\frac{2m_p q_s}{(r^2+a^2)^{1/2} dt/d\lambda}
\frac{\psi^{s,\pm}_{\ell mkn} S^{s=0}_{\ell mkn}}{W}\Bigg]_{r=r(t),\theta=\theta(t)}\label{eq:Zs:ampl}
\end{equation}
The functions $I^{s,\pm}[r(t),\theta(t)]$ are recast as the Fourier series 
\begin{equation}
I^{s,\pm}(t) = \sum_{kn} I^{s,\pm}_{kn} e^{-in\Omega_r t}
e^{-ik\Omega_\theta t} \;.
\end{equation}
Because the coefficients $I^{s,\pm}_{kn}$ depend on the radial and polar coordinates $r= r(w_r\equiv \Upsilon_r\lambda),\theta=\theta(w_\theta\equiv\Upsilon_\theta\lambda)$,
they can be transformed as functions of angles $(\varphi,\chi)$
using Eq.~\eqref{eq:rpsi} and Eq.~\eqref{eq:zchi}.
The coefficients can be integrated over $(\varphi,\chi)$
\begin{equation}
\begin{aligned}    
Z^{s,\pm}_{\ell mkm} =& \frac{T_r}{2\pi}\frac{T_\theta}{2\pi}
\int_0^{T_r}\int_0^{T_\theta} I^{s,\pm}(t) 
e^{-in\Omega_r t} e^{-ik\Omega_\theta t} dt \\
=& \frac{1}{2\pi\Gamma}\int_0^{2\pi} d\chi \int_0^{2\pi} d\varphi
\frac{dw_r}{d\varphi}\frac{dw_\theta}{d\chi}
I^{s,\pm}(\varphi,\chi) 
\\& \times e^{i[\omega_{mkn}t(\varphi,\chi) - m\phi(\varphi,\chi)]}\;,
\end{aligned}
\end{equation}
two quantities can be integrated with the trapezoid formula after setting a tolerable error.

\section{Sensitivity curve of LISA}\label{appC}
The sky-averaged detector sensitivity for LISA can be give by in \cite{LISA:2017pwj,Babak:2017tow}
\begin{eqnarray}
	S_n(f)&=&\frac{20}{3}\frac{4S_{n}^\mathrm{acc}(f)+2S_{n}^\mathrm{loc}+S_{n}^\mathrm{sn}+S_{n}^\mathrm{omn}}{L^2} \nonumber \\
	& & \times
	\left[1+\left(\frac{2Lf}{0.41 c}\right)^2\right],
	\label{eq:sens}
\end{eqnarray}
where $L=2.5\times 10^{9}m$ is the arm length among satellites, and the noise 
$S_{n}^\mathrm{acc}(f)$, $S_{n}^\mathrm{loc}$,
$S_{n}^\mathrm{sn}$ and $S_{n}^\mathrm{omn}$ result from the low-frequency acceleration, local interferometer noise,
shot noise and other measurement noise, respectively. 
They can be written as the following according to LISA Pathfinder~\cite{Armano:2016bkm}
\begin{eqnarray}
	S_{n}^\mathrm{acc}(f) & = & \left\{9 \times 10^{-30}+3.24 \times 10^{-28}\left[\left(\frac{3\times10^{-5}~\mathrm{Hz}}{f}\right)^{10} \right. \right. \nonumber \\
	&  & \left. \left. + \left(\frac{10^{-4}~\mathrm{Hz}}{f}\right)^{2}\right]\right\}\frac{1}{(2\pi f)^4}\,\mathrm{{m^2\,Hz}^{-1}},
\end{eqnarray}
and the other noise expression are of the following
\begin{equation}
	\begin{split}
		&S_{n}^\mathrm{sn}= 7.92\times10^{-23}~\mathrm{{m}^2\,{Hz}^{-1}},\\
		&S_{n}^\mathrm{omn}=4.00\times10^{-24}~\mathrm{{m}^2\,{Hz}^{-1}},\\
		&S_{n}^\mathrm{loc}= 2.89\times10^{-24}~\mathrm{{m}^2\,{Hz}^{-1}}.
	\end{split}
\end{equation}

\section{Comparison of fluxes with previous works}\label{flx comp}
In this section, we plan to examine the numerical fluxes from our code by comparing the energy and angular momentum fluxes for
the scalar and gravitational radiation in previous works~\cite{Warburton:2010eq,Glampedakis:2002ya,Drasco:2005kz}. 
We present the comparison of fluxes from our code and previous papers, incorporating the relative difference of fluxes for several orbital setting in Table~\ref{tab:fluxes:compare}.

\begin{widetext}

\begin{table}[htbp!]
\centering
\begin{tabular}{cccc|ccccccccc}
\hline
\hline
$a$ &   $p$ & $e$   &$x$ &  $\dot{E}^{\text{prev}}_\text{S}$ & $\dot{E}_\text{S}$ & $\text{Rel. Diff.}$ & $\dot{L}^{\text{prev}}_\text{S}$ & $\dot{L}_\text{S}$ & $\text{Rel. Diff.}$ 
& $\dot{Q}^{\text{prev},\infty}_\text{S}$ & $\dot{Q}_\text{S}$ & $\text{Rel. Diff.}$
\\
\hline
0.9  &10  &0.2  &0  & $2.686\text{e-5}$&  $2.679\text{e-5}$ & 2.71\% & $8.359\text{e-4}$ & $8.358\text{e-4}$ & 0.13\%
& $-$ & $-$   & $-$
\\
0.9  &10  &0.5  &0  & $2.468\text{e-5}$&  $2.471\text{e-5}$ & 1.25\% & $6.296\text{e-5}$ & $6.294\text{e-4}$ & 0.33\%
& $-$ & $-$   & $-$
\\
\hline \hline
$a$ &   $p$ & $e$   &$x$ &  $\dot{E}^{\text{prev},\infty}_\text{G}$ & $\dot{E}_\text{G}$ & $\text{Rel. Diff.}$ & $\dot{L}^{\text{prev},\infty}_\text{G}$ & $\dot{L}_\text{G}$ & $\text{Rel. Diff.}$
& $\dot{Q}^{\text{prev}}_\text{G}$ & $\dot{Q}_\text{G}$ 
& $\text{Rel. Diff.}$
\\
0.9  &6  &0.3  &0.94 & $6.80432\text{e-5}$  &$6.80421\text{e-5}$ & $1.61\text{e-3}\%$  & $8.62437\text{e-3}$  & $8.62428\text{e-3}$  & $1.04\text{e-3}\%$
& $6.11706\text{e-3}$  & $6.11758\text{e-3}$ & $8.51\text{e-3}\%$
\\
0.9  &6  &0.3  &0.5 & $8.31504\text{e-4}$  &$8.31495\text{e-4}$ & $1.08\text{e-3}\%$  & $6.48662\text{e-3}$  & $6.48582\text{e-3}$  & $1.23\text{e-2}\%$ 
& $5.86987\text{e-3}$  & $6.11758\text{e-3}$ & $2.33\text{e-2}\%$
\\
\hline
\hline
\end{tabular}
\caption{Relative percentage difference of fluxes from previous
papers and values computed in this work. The spin of the primary is fixed $a=0.9$. The superscript \text{``prev"} represents the fluxes of ~\cite{Warburton:2010eq,Drasco:2005kz}.
Note that the energy flux has an unit of $(m_p/M)^2$, the angular momentum flux carries the unit of $m_p^2/M$ and the Carter constant
flux has the unit of $(m_p^2 M)$.
}\label{tab:fluxes:compare}
\end{table}
\end{widetext}


\makeatletter
\bibliographystyle{apsrev}
\bibliography{JN1}
 
\end{document}